\documentstyle[12pt]{article}

\input{psfig.tex}

\def \div{{\rm div}}
\def \cha{\widehat}

\begin{document}

\title{Exact solutions of the aspherical axisymmetric gravo-magnetic condensation}

\author{Patrick Hennebelle \\
 Laboratoire de radioastronomie millim{\'e}trique, 
 UMR 8540 du CNRS, \\
{\'E}cole normale sup{\'e}rieure,
 24 rue Lhomond, 75231 Paris cedex 05,
France \\
and \\
 Department of Physics and Astronomy,Cardiff University, \\
 PO Box 913, 5 The Parade, Cardiff CF24 3YB, Wales, UK
}

\maketitle


\abstract{ 
We find exact and explicit self-similar solutions of 
the axisymmetric MHD
equations  of a self-gravitating polytropic gas. 
These solutions are able to describe a  
 flat (uniform density) subsonic internal  core,    
contracting homologously, of a collapsing  cloud with
aspherical pressure,
 aspherical gravitational potential, magnetic
field and rotation. Two of the solutions describing the collapse
 of a rotating and magnetized cloud present  outflow at the pole 
or at the equator.
}

\section{Introduction}

\subsection{Previous work}
The gravitational contraction of a  cloud has been extensively
studied numerically and analytically in the context of star formation 
and molecular cloud core collapse. Most of the analytical studies assume a
polytropic equation of state: $D/Dt ( P \rho ^{-\gamma} ) =0$, a simple,
spherical, cylindrical or disk-like geometry, and look for exact or approximate
solutions using various analytical techniques (e.g. self-similarity).

\subsubsection{Hydrodynamical collapse}
 Most of the available work adresses the problem
of hydrodynamical spherical collapse (Penston 1966, 1969a, 1969b,
 Larson 1969, Shu 1977, Hunter 1977, 
Goldreich \& Weber 1980, Yahil 1983, Munier \& Feix 1983, Bouquet et
al. 1985, Whitworth \& Summers 1985) using self-similar approaches.

Larson (1969) and Penston (1969) first reduced the fluid equations
 of a self-gravitating
isothermal gas into ordinary differential equations of the radius, $r$ only. 
They also 
derived an exact solution, that presents uniform density and homologous 
velocity in the internal part of the cloud (see Fig.~\ref{larson-penston}).
 At some point, the velocity (relatively to the similarity profile)
 reaches the sound speed and a bifurcation
occurs, the density decreases and the solution 
tends asymptotically to the singular isothermal 
sphere. This result has been demonstrated analytically by Whitworth \&
Summers (1985) in the isothermal case ($\gamma=1$) and by 
 Bouquet et al. (1985) for any value of $\gamma$. 
Whitworth \& Summers find that the central density is quantised, i.e.
there are bands of central density for which the solution does not
go through the sonic surface. They claim that the Larson-Penston solution,
 that belongs to the first allowed band, represents the subsonic
core of a collapsing cloud and is able to describe better the gravitational 
collapse than other solutions found by Hunter (1977) and Shu (1977) that
belong to the other bands and that are much more centrally peaked.

These studies leave aside important aspects 
such as rotation or magnetic field (Mouschovias 1978). 

\subsubsection{MHD collapse}
Some more recent works  investigate the collapse of a
magnetized gas, making various approximations. Galli \& Shu (1993) use
a self-similar approach and  develop a perturbative solution,
Nakamura et al. (1999), Contopoulos et al. (1998) and Basu (1997)
 consider the thin disk approximation, 
 Chiueh \& Chou (1994) and Li (1998a, 1998b) neglect the magnetic
tension and assume an isotropic magnetic support. A notable
exception is the work of Low (1992) who was able, using a self-similar
analysis, to find exact
solutions of the collapse of the $\gamma = 4/3 $ magnetized polytrope. 
Low reduces the MHD equations to the static equations modified
by a radial static force, that is to say, the Low solutions  
  are static in the similarity space. 
An important restriction of his work is that 
there is an equilibrium between the non radial components of the
 Lorentz, the gravitational and the pressure forces and consequently
the
orthoradial and azimuthal components of the velocity field vanish.
  Thus, as far as the author knows, and in spite of the importance of considering 
 these processes together, there are no exact solutions in the
literature  describing a
gravitational collapse of a  magnetized  and rotating cloud. 

\subsection{Aim of the paper} 
In this paper, we find  exact and explicit solutions describing
 the gravitational collapse 
or expansion of  self-gravitating polytropic 
gas of uniform density   including 
 magnetic field, rotation,
aspherical thermal pressure and aspherical gravitational potential. 
The main difficulty and originality of the present study
  is that the various fields have  a non trivial  angular dependence. 

The derivation of exact solutions  is always of great
interest, they 
 give  explicit examples of how the different terms
counterbalance and  can constitute the starting point of further
analytical studies like stability
or bifurcation (Goldreich \& Weber 1980, Bouquet el al. 1985, 
Blottiau et al. 1988, Ringeval \& Bouquet 2000).
Such solutions can also be used  as benchmarks,  crucial in the testing of
the  complex MHD codes developed for studies related to the
gravitational collapse. Indeed, the importance of code testing cannot
be over emphasized.

In Sect.~2, we apply the dynamical rescaling method introduced by Munier \&
Feix (1983), to the  MHD equation of a self-gravitating
polytropic gas. Considering  axisymmetric geometry, density 
independent of $r$ and homologous velocity fields, we first
 reduce the system  into a set of 8  equations that depend  on
 $\cha{t}$, the rescaled time, and $\theta$ the colatitude only 
and finally reduce it into 7 ordinary differential equations of $\theta$.

In Sect.~3, we restrict the problem further considering a uniform
density field and homologous velocity. 
We find various exact solutions of the stationary (with respect
to the rescaled time, $\cha{t}$) hydrodynamical and MHD equations.
Some of the solutions with rotation and magnetic field have 
radial velocity fields that become positive (outflow)
 at the pole or at the equator.  
 Sect.~4  presents various  solutions, able to describe a collapse,
  with an anisotropic thermal pressure.
Depending on $\gamma$, the polytropic index, some of the solutions present
outflows at the pole or at the equator.
 Sect.~5 concludes the paper. 

\section{Reduction of the system}

\subsection{The equations}

We consider the ideal MHD equations, 
 in spherical coordinates,  of a polytropic self-gravitating gas.
 We thus assume that the magnetic field is
perfectly coupled to the gas, i.e. we do not take into account
ambipolar diffusion.    
In the  usual notation, we have:

\begin{eqnarray}
\partial _t \rho + {\rm div} (\rho {\bf V}) =0
\label{consmat}
\end{eqnarray}


\begin{eqnarray}
\nonumber
\rho \left( \partial _t {\bf V} + {\bf V} .{\rm {\bf grad}} \,{\bf V} \right) 
&=& \\
 - {\rm {\bf grad }} \, P &+& \rho \, {\rm  {\bf grad}} \Phi + {1 \over \mu _0} 
{\rm {\bf rot}} {\bf B} \land {\bf B}
\label{consmom}
\end{eqnarray}

\begin{eqnarray}
 \partial _t \left( P \rho ^{-\gamma} \right)
+ {\bf V} . {\rm {\bf grad}} (P \rho ^{-\gamma}) =0
\label{consener}
\end{eqnarray}


\begin{eqnarray}
\Delta \Phi = - 4 \pi G \,  \rho,
\label{eqpoisson}
\end{eqnarray}


\begin{eqnarray}
\div {\bf B} = 0,
\label{eqdivb}
\end{eqnarray}

\begin{eqnarray}
\partial _t {\bf B} + {\rm {\bf rot}} ( {\bf V} \land {\bf B}) = {\bf 0}
\label{eqmag}
\end{eqnarray}




\subsection{Magnetic field}
In the following, we consider  axisymmetric geometries
 ($\partial _\phi = 0$) and  use the divergence free form for
the magnetic field (Low 1992):
\begin{eqnarray}
B_r = - {1 \over r ^2 \sin \theta} 
\partial _\theta H _1(t,r,\theta),
\label{eqpot1}
\end{eqnarray}

\begin{eqnarray}
B_\theta =  {1 \over r  \sin \theta} 
\partial _ r H _1(t,r,\theta),
\label{eqpot2}
\end{eqnarray}

\begin{eqnarray}
B_\phi =  {1 \over r ^2 } H _2(t,r,\theta).
\label{eqpot3}
\end{eqnarray}
where $H _2$ has been introduced in order to have the same
 physical dimension as $H _1$. 

Eq.~\ref{eqdivb} is automatically
satisfied and Eq.~\ref{eqmag} reduces to the two equations:

\begin{eqnarray}
\partial _t H _1 + V_r \partial _r H _1 +
 {1 \over r} V_\theta \partial _\theta H _1 = 0, 
\label{eqmaga1}
\end{eqnarray}

\begin{eqnarray}
\nonumber
\partial _t H _2 + r \partial _r 
\left( {1 \over r \sin \theta} V _\phi 
\partial _\theta H _1 + {1 \over r } V_r H _2  \right) \\
 + { 1 \over r} \partial _\theta \left( V_\theta H _2
 - {r \over \sin \theta} V _\phi \partial _r H _1  \right) = 0.
\label{eqmaga2}
\end{eqnarray}

\subsection{Dynamical rescaling}
The study of the system of Eqs.~\ref{consmat}-\ref{eqmag} is carried
out using a method which generalizes the usual self-similar technique. 
Following Munier \& Feix (1983) and Bouquet et al. (1985), 
we perform the following transformation:
\begin{eqnarray}
\nonumber
dt & = & \tau ^2(t) \, d \cha{t}, \\
\nonumber
r & = & a(t) \, \cha{r}, \\
\nonumber
P (t,r,\theta) & = & \pi (t) \, \cha{P}(\cha{t},\cha{r},\theta), \\
\label{transfo}
\rho (t,r,\theta) & = & d(t) \, \cha{\rho}(\cha{t},\cha{r},\theta), \\
\nonumber
\Phi (t,r,\theta) & = & f(t) \, \cha{\Phi}(\cha{t},\cha{r},\theta), \\
\nonumber
 H _1 (t,r,\theta) & = & h(t) \, \cha{H} _1(\cha{t},\cha{r},\theta), \\
 H _2 (t,r,\theta) & = & h(t) \, \cha{H }_2(\cha{t},\cha{r},\theta). 
\nonumber
\end{eqnarray}

This procedure generalizes self-similarity since, the transformed fields 
(e.g. $\cha{\rho}$) are allowed to depend both on rescaled space 
and time whereas the
self-similar solutions depend on rescaled space only.  
The time dependence of the new fields allows (at least numerically) the study 
of the convergence toward the self-similar solutions. This procedure
 also allows the study
of the dynamical stability  of the self-similar solution in the rescaled space.

The derivative against time and space are given by:
\begin{eqnarray}
\nonumber
\partial _t &=& {1 \over \tau ^2} \partial _{\cha{t}} - { \dot{a} \over a }
\cha{r} \partial _ {\cha{r}} \\
\partial _r &=& {1 \over a} \partial _{\cha{r}}
\end{eqnarray}

With :
\begin{eqnarray}
\nonumber
\cha{V} _r & = & { \partial  \cha{r} \over \partial \cha{t}} , \\
\label{vitesse_def}
\cha{V} _\theta & = & \cha{r} {\partial \theta \over \partial \cha{t} } , \\
\cha{V} _\phi & = & \cha{r} \sin \theta {\partial \phi \over \partial \cha{t} }  ,
\nonumber
\end{eqnarray}
we have:
\begin{eqnarray}
\nonumber
V _r & = & { \partial  r \over \partial t}  = { \partial  a(t) \cha{r}
\over \partial t} 
= \dot{a}(t) \cha{r} + {a \over \tau ^2}  \cha{V  }_r, \\
\label{vitesse}
V _\theta & = & r {\partial \theta \over \partial t} =
  {a \over \tau ^2}  \cha{V  }_\theta, \\
  V _\phi & = & r \sin \theta {\partial \phi \over \partial t} =
  {a \over \tau ^2}  \cha{V  }_\phi,
\nonumber
\end{eqnarray}

After some algebra, one finds that the system of 
Eqs.~\ref{consmat}, \ref{consmom},
 \ref{consener}, \ref{eqpoisson}, \ref{eqmaga1},
 \ref{eqmaga2} becomes in spherical coordinates 
(as the study of the angular dependence of the fields  is
our  main goal, we write  these equations explicitly):

\begin{eqnarray}
\nonumber
\partial _ {\cha{t}} \cha{\rho} + {1 \over \cha{r} ^2} \partial _{\cha{r}} 
 \left(\cha{r} ^2 \,  \cha{V}_r \, \cha{\rho} \right)     
+ {1 \over \cha{r} \sin \theta } \partial _ \theta 
\left(\sin \theta \, \cha{V}_ \theta \, \cha{\rho} \right) = \\
   - \tau ^2 \left( { \dot{d} \over d}  +  3 { \dot{a} \over a}  \right) \cha{\rho},
\label{t_consmat}
\end{eqnarray}

\begin{eqnarray}
\nonumber
  \partial _{\cha{t}} \cha{V}_r + \cha{V}_r 
\partial _ {\cha{r}} \cha{V}_r
+ {1 \over \cha{r} } \cha{V}_\theta \partial _\theta \cha{V}_r 
 - { \cha{V}_\theta ^2 + \cha{V} _\phi ^2 \over \cha{r}} 
  + \\ 
 2 \tau ^2 \left( { \dot{a} \over a} - { \dot{\tau}   \over \tau } \right) \cha{V} _r 
+ \tau ^4 {\ddot{a} \over a} \cha{r} = 
   - {\pi \tau ^4 \over a ^2 d } { \partial _ {\cha{r}} \cha{P} \over \cha{\rho} } 
+ 
{f \tau ^4 \over a ^2}  \partial _{\cha{r}} \cha{\Phi} -  
\label{t_consmom1} \\ 
{h^2  \tau ^4 \over a ^6 d} {1 \over \cha{\rho} } 
{1 \over \mu _0 \cha{r}^2}  \left( 
 { 1 \over  \cha{r} }
 \cha{H} _2   
 \partial _{\cha{r}} \left( {\cha{H}_2 \over \cha{r}} \right)  +
 \right. { 1 \over  \sin \theta} 
\partial_{\cha{r}} \cha{H}_1 \times  \nonumber \\
 \left.
 \left( \partial _{\cha{r}} 
\left( {1 \over \sin \theta} \partial _{\cha{r}} \cha{H}_1 \right)
  + \partial _\theta \left( {1 \over \cha{r}^2 \sin \theta } 
\partial _\theta \cha{H}_1  \right)  \right)  \right),
\nonumber
\end{eqnarray}

\begin{eqnarray}
\nonumber
  \partial _{\cha{t}} \cha{V}_\theta + 
\cha{V}_r \partial _{\cha{r}} \cha{V}_\theta
+ {1 \over \cha{r} } \cha{V}_\theta \partial _\theta \cha{V}_\theta 
 + { \cha{V} _r \cha{V}_\theta   \over \cha{r}} -
 { \cha{V} _\phi ^2 \cot \theta \over \cha{r}} \\
+ 2 \tau ^2 \left( { \dot{a} \over a} - { \dot{\tau}   \over \tau }
\right) \cha{V} _\theta  = 
   - {\pi \tau ^4 \over a ^2 d} {1 \over \cha{r}  } 
{ \partial _\theta \cha{P} \over \cha{\rho} } 
+ {f \tau ^4 \over a ^2 } {1 \over \cha{r} }  \partial _\theta
\cha{\Phi}  - \nonumber \\
{h^2 \tau ^4 \over a ^6 d} {1 \over \cha{\rho} }
 { 1 \over \mu _0  \cha{r}^3 \sin \theta} \left(
 \cha{H}_2 
 \partial _ \theta \left( {\sin \theta \over \cha{r}^2 } \cha{H}_2 \right)
 +
\right.
\label{t_consmom2} \\
\left.
 \partial _\theta \cha{H} _1 \left( \partial _{\cha{r}}
 \left( {1 \over  \sin \theta} \partial _{\cha{r}} \cha{H}_1 \right) + 
 \partial _\theta  \left( 
{ 1 \over \cha{r}^2 \sin \theta } \partial _\theta \cha{H}_1 \right)
 \right)
 \right),
\nonumber
\end{eqnarray}

\begin{eqnarray}
\nonumber
  \partial _{\cha{t}} \cha{V}_\phi + \cha{V}_r \partial
_{\cha{r}} \cha{V}_\phi
+ {1 \over \cha{r} } \cha{V}_\theta \partial _\theta \cha{V}_\phi 
 + { \cha{V} _r \cha{V}_\phi   \over \cha{r}} + {\cha{V} _\theta
\cha{V} _\phi  \cot \theta \over \cha{r} } \\ \label{t_consmom3}
 + 2 \tau ^2 \left( { \dot{a} \over a} - { \dot{\tau} \over \tau}
\right) \cha{V} _\phi = 
{h^2 \tau ^4 \over a ^6 d}
{1 \over \cha{\rho}}  {1 \over \mu _0 \cha{r}^2 \sin \theta}
\times \\
 \left( { 1 \over  \sin \theta  }
 \partial _{\cha{r}} \cha{H} _1  
\partial _\theta \left( {\sin \theta \over \cha{r}^2 } \cha{H} _2 
\right)  -
{ 1 \over \cha{r}  } \partial _\theta \cha{H} _1 
 \partial _ {\cha{r}} \left( {1 \over \cha{r} } \cha{H} _2 \right) \right),   \,
\nonumber
\end{eqnarray}

\begin{eqnarray}
\nonumber
  \partial _{\cha{t}} \left( \cha{P} \cha{\rho} ^{-\gamma} \right)
 + \cha{V} _r \partial _{\cha{r}} \left( \cha{P} \cha{\rho} ^{-\gamma} \right)
+ {1 \over \cha{r} } \cha{V}_\theta \partial _\theta 
\left( \cha{P} \cha{\rho} ^{-\gamma} \right)  =  \\
- \tau ^ 2{ d _t (\pi d^{-\gamma}) \over \pi d^{-\gamma} }
 \cha{P} \cha{\rho} ^{-\gamma},
\label{t_consener}
\end{eqnarray}

\begin{eqnarray}
{ 1 \over \cha{r}} \partial ^2 _{\cha{r}^2} (\cha{r} \cha{\Phi}) +
{1 \over \cha{r}^2 \sin \theta} \partial _\theta 
\left( \sin \theta \,  \partial _\theta \cha{\Phi} \right) 
  = - 4 \pi G \,  { a ^2 d \over f } \cha{\rho},
\label{t_eqpoisson}
\end{eqnarray}

\begin{eqnarray}
\partial _{\cha{t}} \cha{H} _1 + \cha{V}_r 
\partial _ {\cha{r}} \cha{H} _1 +
 {1 \over \cha{r}} \cha{V}_\theta \partial _\theta \cha{H} _1 
 = - \tau ^2 { \dot{h} \over h} \cha{H} _1,
\label{t_eqmaga1}
\end{eqnarray}

\begin{eqnarray}
\partial _{\cha{t}} \cha{H} _2 + \cha{r} \partial _{\cha{r}} 
\left( {1 \over \cha{r} \sin \theta} \cha{V} _\phi 
\partial _\theta \cha{H} _1 + {1 \over \cha{r} } \cha{V}_r \cha{H} _2  \right)
\nonumber \\
 + { 1 \over \cha{r}} \partial _\theta \left( \cha{V}_\theta \cha{H} _2
 - {\cha{r} \over \sin \theta} \cha{V} _\phi \partial _{\cha{r}}
\cha{H} _1  \right)  =  - \tau ^2 { \dot{h} \over h} \cha{H} _2.
\label{t_eqmaga2}
\end{eqnarray}

We choose the scaling in Eqs.~\ref{transfo} in order to cancel explicit 
time dependencies in system of Eqs.~\ref{t_consmat}-\ref{t_eqmaga2}. 
Thus we impose that:
\begin{eqnarray}
\left\{
\begin{array} {l}
\tau ^2 \left(  \dot{d} / d  +  3  \dot{a} / a  \right) =
K_1 \; , \; 
\tau ^2 \left(  \dot{a} / a -  \dot{\tau}   / \tau 
\right) = K _2  \;  ,  \;  \\ 
\tau ^4 \ddot{a} / a = K_3  \; ,  \;   
\pi \tau ^4 / (a ^2 d)  = K_4  \; ,  \; \\  
f \tau ^4 / a ^2 = K_5  \; ,  \;  
h^2  \tau ^4 / a ^6 d = K_6  \;  ,  \; \\
a ^2 d / f  = K _7 \; , \;   
\tau ^ 2  (d / dt (\pi d^{-\gamma})) / \pi d^{-\gamma}  = K _8
 \; ,  \; \\  
\tau ^2  \dot{h} / h = K _9 \; , 
\end{array}
\right.
\label{transfo_cond}
\end{eqnarray}
where $K _i$ are real numbers. 
Although there are  6 variables only and nine constraints, the
 system of Eqs.~\ref{transfo_cond}
admits the following solution:
\begin{eqnarray}
\left\{
\begin{array} {l}
\tau(t) = \sqrt{\Omega _0  \omega t + 1} \; , \;  
a(t) = \left( { \Omega _0 \omega t + 1} \right)
 ^{\lambda} \; , \; \\ 
d(t) = \left( \Omega _0 \omega t + 1 \right)^{-2} \; , 
\pi (t) = ( \Omega _0 \omega t + 1) ^{2 \lambda -4} \; , \; \\
f(t) = ( \Omega _0 \omega t + 1) ^{2 \lambda -2} \; , \;
h (t) = ( \Omega _0 \omega t + 1) ^{3 \lambda -2}.
\end{array}
\right.
\label{transfo_sol}
\end{eqnarray}
where $\omega$ and  $\lambda$  are two real numbers,
$\rho _0$ is the density at origin and the Jeans frequency:  
$\Omega _0 = \left(4 \pi G \rho _0 \right)^{1/2}$ has been introduced
 for later simplification.

With Eqs.~\ref{transfo_cond}, one obtains the dependence of the nine constants
  $K _i$  on $\omega$ and $\lambda$:
\begin{eqnarray}
\left\{
\begin{array} {l}
K_1 = \Omega _0 \omega ( 3 \lambda -2) \; , 
\; K_2 = \Omega _0 \omega (\lambda - 1/2) \; , \; \\
K_3 = (\Omega _0 \omega) ^2 \lambda ( \lambda -1)  \; , \; K_4 =1 \; , \; \\
 K_5=1 \; , \; K_6=1 \; , \; K_7=1 \; , \; \\
K _8 = 2 \Omega _0 \omega ( \lambda + \gamma -2  ) \; , \;
 K _9 = \Omega _0 \omega (3 \lambda -2 ). 
\end{array}
\right.
\label{transfo_res}
\end{eqnarray}

With Eqs.~\ref{transfo_cond} and~\ref{transfo_res}, the
system of Eqs.~\ref{t_consmat}-\ref{t_eqmaga2} is  independent of the
variable $t$. It can be seen that due to the transformation stated by
Eqs.~\ref{transfo} new terms, that can be seen e.g. as 
source terms, appear in the new frame defined by the variable
$\cha{t}$ and $\cha{r}$ (see e.g. Bouquet et al. 1985
for a discussion).

If $\omega$ is positive, the transformation is defined for any value
of $t$ and describes an expansion, and if $\omega$ is negative the
density becomes infinite at $t=-1/(\Omega _0 \omega)$ and 
the transformation describes a collapse.

The transformation performed in this paper is slightly more general
than the transformation performed by previous authors (Munier \& Feix
1983, Bouquet et al. 1985) who assume that $\pi (t) = d ^{\gamma}
(t)$, i.e. they impose that energy is conserved in the similarity space.
 This choice leads to $\lambda = 2 - \gamma$, whereas in the present study
 $\lambda$ is a free parameter, energy is not conserved in the similarity
space since a source term appears in Eq.~\ref{t_consener}.  

\subsection{Reduction into systems of  ordinary differential equations}
In order to reduce further Eqs.~\ref{t_consmat}-\ref{t_eqmaga2},
it is worthwhile to reduce these equations into  systems of ordinary 
differential equations.

\subsubsection{Ordinary equations of the radius}
The first and obvious possible reduction is to consider fields independent
on $\theta$. In this case, $V _\theta$, $V _\phi$, $H _1$, $H _2$ must 
vanish and with $\lambda=2-\gamma$, one obtains the following equations: 
\begin{eqnarray}
\partial _ {\cha{t}} \cha{\rho} + {1 \over \cha{r} ^2} \partial _{\cha{r}} 
 \left(\cha{r} ^2 \,  \cha{V}_r \, \cha{\rho} \right)   = 
   -   \Omega _0 \omega  (4 - 3 \gamma) \cha{\rho},
\label{t_consmat_r}
\end{eqnarray}

\begin{eqnarray}
\nonumber
  \partial _{\cha{t}} \cha{V}_r + \cha{V}_r 
\partial _ {\cha{r}} \cha{V}_r  +  
 \Omega _0 \omega (3 - 2 \gamma ) \cha{V} _r 
&+& (\Omega _0 \omega)^2 (2-\gamma) (1-\gamma) \cha{r}  \\ =  
   &-&  { \partial _ {\cha{r}} \cha{P} \over \cha{\rho} } 
+   \partial _{\cha{r}} \cha{\Phi},   
\label{t_consmom_r}  
\end{eqnarray}

\begin{eqnarray}
{ 1 \over \cha{r}} \partial ^2 _{\cha{r}^2} (\cha{r} \cha{\Phi}) 
  = - 4 \pi G \,   \cha{\rho}.
\label{t_eqpoisson_r}
\end{eqnarray}
 which is  the system obtained
by Yahil (1983) and by Bouquet et al. (1985).
In the isothermal case ($\gamma =1$) it reduces to 
the system obtained by Penston (1969) and Larson (1969)
 and extensively studied by Shu (1977), Hunter(1977) and 
Whitworth \& Summers (1985). 

\subsubsection{Ordinary equations of the colatitude}
Another possibility is to reduce the system of 
Eqs.~\ref{t_consmat}-\ref{t_eqmaga2}, into a system that depends on
the angle $\theta$ only.

This reduction is obtained by considering the fields:

\begin{eqnarray}
\left\{
\begin{array} {l}
\cha{t} = \widetilde{t}  / \Omega _0 \; , \; 
\cha{V} _r (\cha{t},\cha{r},\theta)  = 
\Omega _0 (V (\widetilde{t},\theta) - \lambda \omega) \cha{r} \; , \; \\
\cha{V} _\theta (\cha{t},\cha{r},\theta)  = \Omega _0 U (\widetilde{t},\theta) \cha{r} \; ,  \; 
\cha{V} _\phi (\cha{t},\cha{r},\theta) =  
\Omega _0 W(\widetilde{t},\theta) \cha{r} \; , \; \\
 \cha{P}(\cha{t},\cha{r},\theta)   = \rho _0 \Omega _0^2  \Pi  (\widetilde{t},\theta) \cha{r}^2 \; , \;  
\cha{\rho}(\cha{t},\cha{r},\theta)  = \rho_0 R (\widetilde{t},\theta)  \; , \; \\
\cha{\Phi} (\cha{t},\cha{r},\theta)  = \Omega _0 ^2 \phi (\widetilde{t},\theta) \cha{r}^2  \; , \; \\
\cha{H} _1 (\cha{t},\cha{r},\theta) = \sqrt{\mu _0 \rho _0} \Omega _0  h _1 (\widetilde{t},\theta) \cha{r}^3 \; , \; \\
\cha{H} _2 (\cha{t},\cha{r},\theta) = \sqrt{\mu _0 \rho _0} \Omega _0  h _2 (\widetilde{t},\theta) \cha{r}^3 \; . \; 
\end{array}
\right.
\label{espace}
\end{eqnarray}
With the definition of $V(\theta)$ (first equation), it is easily seen
with Eqs.~\ref{vitesse} and~\ref{transfo_sol} that $V _r(r,\theta)
\propto V(\theta)$.  We also have $V _\theta(r,\theta)
\propto U(\theta)$ and $V _\phi (r,\theta)
\propto W(\theta)$.

As for the Larson-Penston solution, 
these fields diverge when $r$ goes to infinity and the solutions are
valid in a finite domain only (until the velocity reaches the velocity
of the sound or magneto-sonic waves). 

 Another important
 restriction is due to the fact that the magnetic field (see
Eqs.~\ref{eqpot1},~\ref{eqpot2} and~\ref{eqpot3}) is proportional to
the radius, $r$, and vanishes at the origin. 
Thus, the magnetic pressure compresses
the gas and enhances the condensation process instead of supporting
the gas against the gravitational collapse. 
However, in the following we will be able to add a uniform field and to avoid this restriction 
for some of the solutions (see Eqs.~\ref{magrad}-\ref{magort}). 

With Eqs.~\ref{transfo_cond},~\ref{transfo_res}
and~\ref{espace},
 the system of Eqs.~\ref{t_consmat}-\ref{t_eqmaga2} becomes:

\begin{eqnarray}
\partial _ {\widetilde{t}} R + 3 V R + {1 \over \sin \theta} 
\partial _\theta \left( \sin \theta U R \right)
= 2 \omega \;  R,
\label{cmat_red}
\end{eqnarray}

\begin{eqnarray}
\nonumber
\partial _ {\widetilde{t}} V +
V ^2  + U  \partial _\theta V - U ^2  - W ^2 
 - \omega \; V    = - 2 { \Pi \over R } 
+ 2 \phi   \\
-{1 \over  R } \left( 2 h _2 ^2  +
 { 18 \over \sin ^ 2 \theta  } h _1 ^ 2   + 
{3 \over \sin \theta}   h_1
\partial _\theta \left( {1 \over \sin \theta }
  \partial _\theta h_1   \right)   \right),
\label{cmom1_red}
\end{eqnarray}

\begin{eqnarray}
\nonumber
\partial _ {\widetilde{t}} U +
2 V U + U \partial _\theta U - W^2  \cot \theta 
 -  \omega \; U  = \\ \nonumber - { \partial _\theta \Pi \over
R } 
+  \partial _\theta \phi \\  - 
{1 \over  R } \left( 
{1 \over \sin \theta } h_2  \partial _ \theta (\sin \theta
h_2)  +  { 6 \over \sin ^2  \theta }  \partial _\theta  h _1 h_1 \right.
\label{cmom2_red} \\
\left.
+ { 1 \over \sin \theta} \partial _\theta h_1 \partial _\theta \left( {1 \over \sin
\theta}  \partial _\theta h _1  \right) \right),
\nonumber
\end{eqnarray}

\begin{eqnarray}
\nonumber
\partial _ {\widetilde{t}} W +
2 V  W + U \partial _\theta W + U  W
\cot \theta    - \omega \; W  = \\
{ 1 \over  R } \left( { 3 \over \sin ^2 \theta } h_1 
\partial _\theta \left( \sin \theta h _2  \right) -
{2 \over \sin \theta}  \partial _\theta h_1 h_2 
  \right),   
\label{cmom3_red}
\end{eqnarray}

\begin{eqnarray}
\partial _ {\widetilde{t}} (\Pi R ^{-\gamma}) +
2 V  \Pi R ^{-\gamma}  
+ U  \partial _\theta 
\left( \Pi R ^{-\gamma}  \right)
= \nonumber \\ - 2 ( \gamma - 2) \omega \; 
 \Pi R ^{-\gamma},   
\label{cen_red}
\end{eqnarray}

\begin{eqnarray}
6 \phi  + { 1 \over \sin \theta } \partial _\theta
\left(  \sin \theta  \partial _\theta \phi  \right)
= - R,
\label{pois_red}
\end{eqnarray}

\begin{eqnarray}
\partial _ {\widetilde{t}} h _1 +
3 V(\theta) h_1 (\theta) + U (\theta)  \partial _\theta h_1(\theta)  =
 2 \omega \;  h _1(\theta),
\label{m1_red}
\end{eqnarray}

\begin{eqnarray}
\partial _ {\widetilde{t}} h_2 +
3 V   h_2 -
  3  \partial _\theta \left( { 1 \over \sin \theta} W \right) h_1  
+ \partial _\theta ( U h _2  )
= 2 \omega \; h _2. 
\label{m2_red}
\end{eqnarray}

The parameter $\lambda$, describing the radius rescaling, 
does not appear any more in these equations
because of the special  dependence on $r$ stated by Eqs.~\ref{espace}
  which indeed is scale invariant. 

This set of equations (Eqs.~\ref{cmat_red}-\ref{m2_red})
 is still quite complex and
 solving it completely is a formidable mathematical
 problem. We will thus look in the next sections for
 particular solutions of the system of 
Eqs.~\ref{cmat_red}-\ref{m2_red}. More precisely, we will look 
 for stationary solutions with respect to the $\widetilde{t}$ variable. 
Such solutions are not  stationary with respect to the physical
time $t$ and in spite of clear physical limitations, 
ought to
describe an aspherical collapse or expansion with complex processes,
like magnetic field or rotation included.

\section{Exact solutions}
In this section, we derive  exact solutions of Eqs.~\ref{cmat_red}-~\ref{m2_red}
stationary with respect to $\widetilde{t}$, (i.e. $\partial _{\widetilde{t}}=0$)
first in the hydrodynamical (i.e. unmagnetized)
case: $h_1=h_2=0$ and then consider the MHD case.
In  appendix A, the solutions are summarized and given explicitly in the
physical space.

\subsection{Uniform density field: gravitational potential and
thermal pressure}
Eqs.~\ref{cmat_red},~\ref{cmom1_red},~\ref{cmom2_red},~\ref{cen_red}, and 
\ref{pois_red}   cannot be easily integrated and in order to further reduce  
these equations,  we look for  solutions with uniform density: 
\begin{eqnarray}
R(\theta)=1.
\label{sol_mat}
\end{eqnarray}

\subsubsection{Gravitational potential}
\label{gravi}

One can easily obtain a solution of Eq.~\ref{pois_red}:
\begin{eqnarray}
\phi (\theta) = - {1 \over 6 } + \phi _1 (3 \cos ^2 \theta -1). 
\label{sol_phy}
\end{eqnarray}

If $\phi _1 \ne 0$, the gravitational potential is aspherical whereas the 
density field is constant.  
This result can be physically understood in the following way.
The fields:
\begin{eqnarray}
\nonumber
\cha{\rho} _\alpha (\cha{r},\theta) &=& {1 \over 4 \pi G} \left( \alpha (\alpha +1) A + \right. \\ 
  && ( \alpha (\alpha +1) -6 ) B \left. \,  (3 \cos ^2 \theta -1) \right) \cha{r} ^ {\alpha-2},  
\label{champ_lim1}
\end{eqnarray}
\begin{eqnarray}
\cha{\Phi} _\alpha (\cha{r},\theta ) = -( A + B (3 \cos ^2 \theta -1) ) 
\cha{r} ^ \alpha. 
\label{champ_lim2}
\end{eqnarray}
are solutions of Eq.~\ref{t_eqpoisson} ($A$ and $B$ are two real numbers). 
The physical requirement $\cha{\rho} _\alpha > 0$ implies that:
\begin{eqnarray}
\begin{array}{l}
\alpha ( \alpha + 1) A + 3 B ( \alpha (\alpha + 1) -6) > 0 \; , \\
 \alpha ( \alpha + 1) A -  B ( \alpha (\alpha + 1) -6) > 0 .
\end{array}
\end{eqnarray}
In the limit $\alpha \rightarrow 2$, it is equivalent to $A >0$.

With these definitions, $ R$ and $ \phi $ 
 are equal to  respectively $  \cha{\rho} _{\alpha \rightarrow 2}  / \rho _0 $
 and $ \cha{\Phi} _ {\alpha \rightarrow  2 }/ (\Omega _0 ^2 \cha{r}
^{2})$.
 In the limit $\alpha
\rightarrow 2$,  the colatitude dependent part of $\cha{\rho} _\alpha$
vanishes whereas this is not the case for $\cha{\Phi} _\alpha$. 
This means, that in the limit $\alpha \rightarrow 2$,
the colatitude dependent part of the density field stated by 
Eqs.~\ref{champ_lim1} and~\ref{champ_lim2} is
dynamically negligible but not gravitationally negligible.  

An explanation for the origin of the aspherical potential
  will be presented in Sect.~\ref{asph_pot}.

If $ \phi _1 < -1/6$ then $\phi(0) < 0$ and $\phi(\pi /2 ) > 0$. The
gravitational force  attracts  towards  the origin
 the gas at the pole  and   moves away  from the origin the gas at the equator.

If  $ \phi _1 > 1/12$ then $\phi(0) > 0$ and $\phi(\pi /2 ) < 0$.
The
gravitational force  attracts  towards  the origin
 the gas at the equator  and  moves away   from the origin the gas at the pole.

\subsubsection{Thermal pressure}
\label{pression thermique}
The thermal pressure given by Eqs.~\ref{espace} is proportional to 
$\Pi(\theta) \cha{r}$, thus a negative value of $\Pi(\theta)$  could appear unphysical 
(negative thermal pressure). However with the assumption stated
 by Eq.~\ref{sol_mat}, it is easily seen that:
\begin{eqnarray}
 P(t,r,\theta) = d(t) ^\gamma P_0 + \pi(t) \cha{P} (\cha{t},\cha{r},\theta),
\label{pression}
\end{eqnarray}
where $P _0$ is a real number, is also a solution of the problem
($d(t) ^\gamma P _0 (d(t) R ) ^ {-\gamma}$ is a constant). 
Thus choosing $P_0 > 0$, allows to have a thermal pressure positive until a
 finite value of $r$.  It  still becomes negative for an
arbitrary high value of $r$, but   with the assumption stated by
Eqs.~\ref{espace}, as we already said, these solutions are  valid only in  a finite
domain. The value of $P_0$ can thus be high enough for the thermal
pressure to be positive in the whole domain of validity.

With the uniform density field assumption, $\cha{P} \ne 0$
implies that the temperature is not a function of the density only
as it is usually assumed. An anisotropic temperature can be due to
anisotropic heating  (Nelson \&
Langer 1997 who consider an anisotropic UV heating), or to
 isotropic heating and  anisotropic optical opacity resulting from the
anisotropic  shape of the cloud (see the discussion of Sect.~\ref{asph_pot}
on the anisotropic sonic surface).

\begin{figure}[htbp]
\centerline{\psfig{file=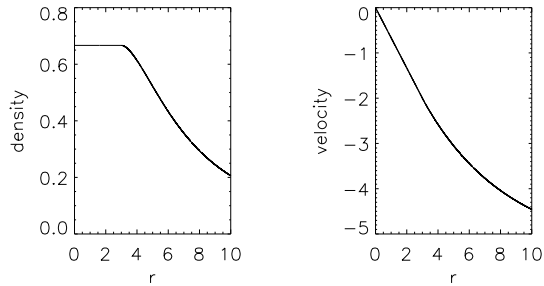,width=8cm}}
\caption{Density and velocity fields corresponding to the Larson-Penston
solution. The density  is constant for $r<3$ and the velocity is 
homologous. A bifurcation occurs at the sonic point located at $r=3$.}
\label{larson-penston}
\end{figure}

In Sect.~3 we will consider only the case where $\cha{P} =0$. 
This implies that the thermal pressure does not have any 
dynamical effect since it is uniform.
 However, as for the Larson-Penston solution, it is expected
that at the sonic point, i.e. when the velocity is equal to the sound speed,
a bifurcation occurs (see Fig.~\ref{larson-penston}). Thus,  thermal pressure  plays an important
r\^ole since it induces the bifurcation and the solutions are not equivalent 
to the solutions describing the collapse of a cold  cosmological cloud that
has no thermal pressure (Lin et al. 1965).

In Sect.~4, we will consider the case where $\cha{P} \ne 0$.
\subsection{Hydrodynamical equations}

\begin{figure}[htbp]
\centerline{\psfig{file=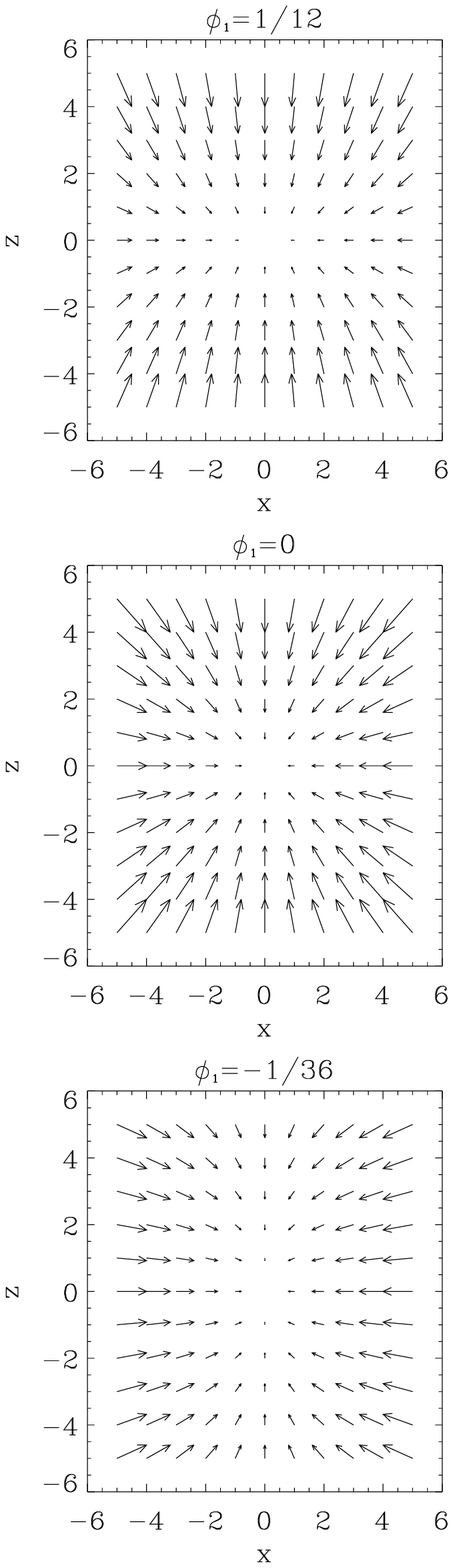,width=8cm}}
\caption{Velocity field  corresponding to collapse with anisotropic 
gravitational potential (solutions stated by
 Eqs.~\ref{sol_hyd1}) with $\phi _1=1/12$ (oblate sonic surface with 
aspect ratio of 2), $0$ (spherical case), $-1/38$ 
(prolate sonic surface with aspect ratio of $5/2$). 
The field is axisymmetric with respect to the z-axis.}
\label{champ_sym}
\end{figure}

In this section, we  consider the aspherical hydrodynamical 
case when the rotation 
is non  zero and  the gravitational 
potential  depends on $\theta$  ($\phi _1 \ne 0$).  
We look  for solutions
with uniform density (Eq.~\ref{sol_mat}) and assume
 that:
\begin{eqnarray}
\begin{array}{l}
V(\theta) = V_0 - U _1 \cos ^2 \theta \; , \;   
U(\theta) =    U _1 \cos \theta \sin \theta \; , \; \\ 
W(\theta) = W _1 \sin  \theta. 
\label{hypo}
\end{array}
\end{eqnarray}

Eqs.~\ref{cmat_red},~\ref{cmom1_red},~\ref{cmom2_red} and \ref{cmom3_red} lead
 to:
\begin{eqnarray}
\left\{
\begin{array} {l}
  U_1 =  3 V_ 0  -2 \omega, \\
V _0 ^2  - W_1 ^2  - \omega V_0 =
   - 1 / 3 - 2 \phi _1, \\
- 2 V_0 U _1 + U _1 ^2 + W_1 ^2 +  
\omega U _1 =   6 \phi _1, \\
W_1 =0 \, {\rm or} \, V _0 =  \omega / 2. 
\end{array}
\right.
\label{cond_mom_rot}
\end{eqnarray}
The first equation derives from the mass conservation whereas the second and the third
 derive from the radial momentum conservation (the orthoradial momentum
conservation leads also to the third equation). 
The term $W _1 ^2$ in the left-hand side of the second and the third
equation is the centrifugal force. The fourth equation derive from 
the azimuthal momentum conservation.

With Eqs.~\ref{cond_mom_rot}, we obtain two
 solutions stated by Eqs.~\ref{sol_hyd1} and~\ref{sol_rot1}.
The first one describes the collapse of an aspherical and non rotating cloud
and the second one the collapse of a rotating and aspherical cloud.

\subsubsection{Spherical solution}

We first consider the spherical case with $U_1=0$, $W _1=0$ and $\phi _1 =0$.
We then have:
\begin{eqnarray}
V _0 = 2/3 \; \omega \; , \; \omega ^2 = 3 / 2,
\label{bouquet_sol}
\end{eqnarray}
which indeed is the solution found by Bouquet et al. (1985) 
and by Penston (1969a) and Larson (1969) in the isothermal case. 

Bouquet et al. (1985) and Whitworth \& Summers (1985), in the case $\gamma=1$, 
 show that at the sonic point, i.e when $\cha{V}=\cha{C}_s$
($\cha{C}_s$ being the sound speed), the solution stated by 
Eq.~\ref{bouquet_sol},
becomes unstable and a bifurcation occurs. After this point, the density
  decreases and one recovers asymptotically the 
Chandrasekhar solutions (Chandrasekhar 1967). 

This is illustrated in Fig.~\ref{larson-penston} that displays
 the density field corresponding
to the  Larson-Penston solution. The bifurcation occurs at $r=3$.

\subsubsection{Aspherical  gravitational potential}
\label{asph_pot}

If $W _1 =0$ (no rotation) and $\phi _1 \ne 0$ 
(anisotropic gravitational field), the solutions of Eqs.~\ref{cond_mom_rot}
are:
\begin{eqnarray}
 \left\{
\begin{array} {l}
\omega = \pm \sqrt{ {3 \over 2} } 
{1 + 12 \phi _1 \over \sqrt{1+24 \phi _1} }  
 \; , \;
 U _1 = - 12 \sqrt { 3 \over 2} {  \phi _1 \over  \sqrt{1 + 24 \phi
_1} } {\omega \over | \omega |} \; , \\
V(\pi /2) = V _0 = \sqrt{ {2 \over 3} } {1 + 6 \phi _1 \over \sqrt{1 + 24 \phi _1}  }
{\omega \over | \omega |}  \; , \; \\
V(0) =  \sqrt{ {2 \over 3} } {1 + 24 \phi _1 \over \sqrt{1 + 24
\phi _1}  }
{\omega \over | \omega |}.
\end{array}
\right. 
\label{sol_hyd1}
\end{eqnarray}
 Due to an aspherical 
(see Eqs.~\ref{sol_mat},~\ref{sol_phy},~\ref{champ_lim1},~\ref{champ_lim2})
  gravitational field  the collapse 
or the expansion are aspherical.
In the limit $\phi _1 \rightarrow 0$, one recovers the
solution stated by Eq.~\ref{bouquet_sol}. A solution exist only if
$\phi _1 > - 1 / 24$.

 It is likely (at least if $\phi _1 \ll 1$)
that as for the Larson-Penston solution a bifurcation
 will occur  at the points of the sonic surface
defined by: $\cha{V}_r ^2 + \cha{V} _\theta ^2 = \cha{C}_s^2$.
With this definition, it is easily seen that the sonic surface
 is an ellipsoid defined by the relation:
\begin{eqnarray}
 { \Omega _0 ^2 V_0 ^2   \over \cha{C}_s ^2} (x^2 + y^2) + 
{\Omega _0 ^2 (V_0 -U_1)^2  \over \cha{C}_s ^2} z^2 = 1,   
\end{eqnarray}
where $x = r \sin \theta \cos \phi$, $y = r \sin \theta \sin \phi$
 and $z = r \cos \theta$.
The length of its  axes are: $a= \cha{C}_s/  (\Omega_0 |V_0|)$ and
 $b= \cha{C}_s / (\Omega_0 |V_0 -U_1| )$,
and the aspect ratio is given by: 
\begin{eqnarray}
{a \over b} = {V_0 - U_1 \over V_0} = { 1 + 24 \phi _1 \over 1 + 6 \phi _1 }.
\end{eqnarray} 
This fact, indeed leads to an explanation for the origin of
 the aspherical gravitational potential since the sonic surface is not
 spherical and since the gas density outside the sonic surface  is not
 uniform.

If $\phi _1 \rightarrow -1/24$ or $\phi _1 \rightarrow \infty$, then
$V(0) \rightarrow \infty$ which means that the sonic surface becomes
smaller and tends to the origin.

We will not consider such bifurcation further even if  this rather
difficult (particularly in two dimensions) question requires an
 accurate and careful investigation at some stage. 

If $\phi _1 > 0$ then $a / b > 1$ and the cloud or at least the 
sonic surface is oblate. It is prolate in the other case. 
The velocity field corresponding to the three values $\phi _1=1/12$
(oblate cloud with aspect ratio equal to 2), $\phi _1=0$ (spherical solution),
 $\phi _1 = -1/36$ (prolate cloud with aspect ratio equal to $5/2$) are
displayed in Fig.~\ref{champ_sym}.

\subsubsection{Rotation} 
We now consider  a non vanishing  azimuthal velocity.

The parameters of the solution  are:
\begin{eqnarray}
\left\{
\begin{array} {l}
\omega = \pm \sqrt{ 4 / 3 + 8 \phi _1 - 4 W_1 ^2} 
\; , \; \\
 \phi _1 = 1 / 12 \; ; \;  U _1 = -  \omega / 2 \; ,  \\
V(\pi/2) = V_0 = \omega / 2 \; , \; V (0) = \omega.
\end{array}
\right. 
\label{sol_rot1}
\end{eqnarray}
The gravitational potential is strongly constrainted: $\phi _1 =
1/12$, and $\phi (0) = 0$. The sonic surface is oblate with an aspect ratio 
of 2, the minor axis being parallel to the rotation axis.
The frequency, $\omega$, decreases with increasing rotation velocity and
the solution becomes stationary for   $W _1 = 1 / \sqrt{2} $, meaning that
for this value the cloud is rotationaly supported.

\subsection{Magnetohydrodynamics}
In this part, we consider  the gravo-magnetic condensation and derive exact solutions.

If $\partial _{\widetilde{t}} = 0$, Eqs.~\ref{cmat_red} and~\ref{m1_red} lead to:
\begin{eqnarray}
K \, R(\theta) U (\theta) \sin \theta = h _1 (\theta),
\label{m1_cond}
\end{eqnarray}
where $K$ is a real number.

We consider first a problem without rotation and 
without magnetic toroidal field.

\subsubsection{Magnetic field without rotation}

\begin{figure}[htbp]
\centerline{\psfig{file=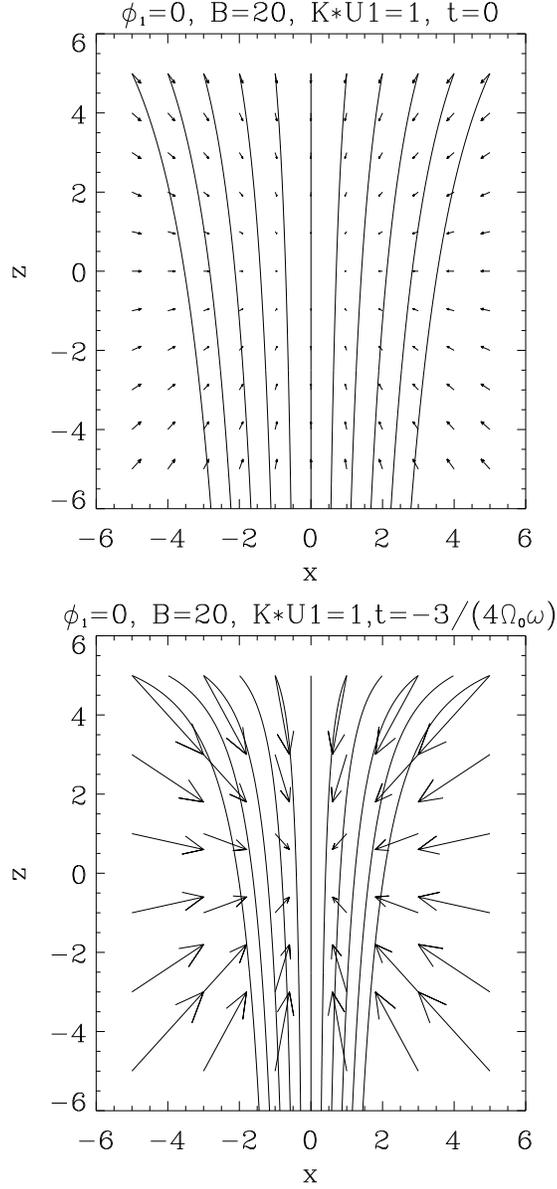,width=8cm}}
\caption{Velocity field and magnetic field lines corresponding
to  Eqs.~\ref{bouquet_sol}  and to
Eqs.~\ref{magrad}-\ref{magort} (with $B=20$ and $K U_1=1$).
The two time steps, $t=0$ and $t=3/(4 \Omega _0 |\omega| )$, are displayed. }
The evolution of the velocity field is self-similar and it is multiplied by 4
between the 2 steps.  The evolution of the magnetic field is not self-similar
because the uniform component and the component stated by Eqs.~\ref{magnetic}
have different temporal evolution.
\label{champ_ev}
\end{figure}

We consider a magnetized, non rotating cloud and
we look for solutions stated by Eqs.~\ref{hypo}. With 
Eq.~\ref{m1_cond}, we have:
\begin{eqnarray}
h _1 (\theta) = K U _1 \, \cos \theta \sin ^2 \theta.
\label{mag_sol}
\end{eqnarray}
The corresponding magnetic components are proportional to:
\begin{eqnarray}
\begin{array}{l}
B _r \propto -( 3 \cos ^ 2 \theta -1 ) \; r,   \\
B _\theta \propto 3 \cos \theta \sin \theta \; r.
\end{array}
\label{magnetic}
\end{eqnarray}

We have:
\begin{eqnarray}
 { 6 \over \sin \theta }  h_1  + \partial _\theta 
\left( {1 \over \sin \theta} \partial _\theta h _1   \right) =0.
\label{m1_nul}
\end{eqnarray}
which  means that ${\bf {\rm {\bf rot}} B = 0} $.
Thus, for the field given by Eq.~\ref{mag_sol} 
the Lorentz force vanishes 
(see Eqs.~\ref{cmom1_red}-\ref{cmom2_red}). 

Before to continue, we demonstrate that adding a spatialy uniform 
magnetic field is possible. 
Let us consider the uniform  magnetic field:
\begin{eqnarray}
\begin{array} {l}
B   _r (t, r ,\theta)= \; \; \; b(t) B \cos \theta   \; , \\
B   _\theta (t, r ,\theta)=  -b(t) B \sin \theta.   
\end{array}
\label{champ_cons}
\end{eqnarray}
where B is a magnetic intensity. Here again the Lorentz force vanishes.
With Eqs.~\ref{eqmag}  and for the velocity field
stated by Eq.~\ref{hypo}, one finds that:
\begin{eqnarray}
b(t) = (1 + \Omega _0 \omega t) ^ {- 2 V _0 / \omega}.
\end{eqnarray}

Consequently,
Eqs.~\ref{consmat}-\ref{eqmag} admit the solutions  given by 
Eqs.~\ref{sol_hyd1} (and by 
Eqs.~\ref{transfo},~\ref{transfo_sol},~\ref{espace},~\ref{hypo})
 with  the magnetic field (since the Lorentz  force associated to the fields stated by 
Eqs.~\ref{magnetic} and~\ref{champ_cons} vanishes):
\begin{eqnarray}
\nonumber
B _r ( t,r, \theta) &=& 
 B  (1 + \Omega _0 \omega t ) ^{- 2 V _0 / \omega}  \cos \theta + \\
\sqrt{\mu _0 \rho _0} &\times& \Omega _0  (1 + \Omega _0
\omega t) ^ {-2}   K ( - 3 U _1 \cos ^2 \theta + U _1  ) \;  r \; , 
\label{magrad}
\end{eqnarray}
\begin{eqnarray}
\nonumber
B _\theta ( t,r, \theta) &=& 
-  B  (1 + \Omega _0 \omega t ) ^{-2 V _0 / \omega}  \sin \theta + \\
\sqrt{\mu _0 \rho _0} &\times& \Omega _0  (1 + \Omega _0
\omega t) ^{-2 } K  (  3 U _1 \cos  \theta \sin \theta  ) \;  r \; ,
\label{magort}
\end{eqnarray}
 for any value of $B$ and $K$.
In the limit $U _1 \rightarrow 0$ (reached for $\phi _1 \rightarrow 0$
in Eqs.~\ref{sol_hyd1} and $\gamma \rightarrow 4/3$ in
Eqs.~\ref{sol_hyd2} and~\ref{sol_hyd3})
and $K \rightarrow \infty$,
 one finds  that they also admit the spherical solutions
(Eqs.~\ref{bouquet_sol}).


Fig.~\ref{champ_ev} displays the velocity field and magnetic field lines
 stated by Eqs.~\ref{bouquet_sol}  with the  magnetic fields
stated by  Eqs.~\ref{magrad}-\ref{magort} ($\phi _1=0$, $B=20$, $K U_1 =1$)
at two times steps ($t=0$ and $t=-3/(4\Omega _0 \omega)$). 
The behaviour of the magnetic field is not 
 self-similar because the two components of Eqs.~\ref{magrad}-\ref{magort}
 are differentially rescaled.


\subsubsection{Rotation and magnetic field} 
\label{rot_mag}

\begin{figure}[htbp]
\centerline{\psfig{file=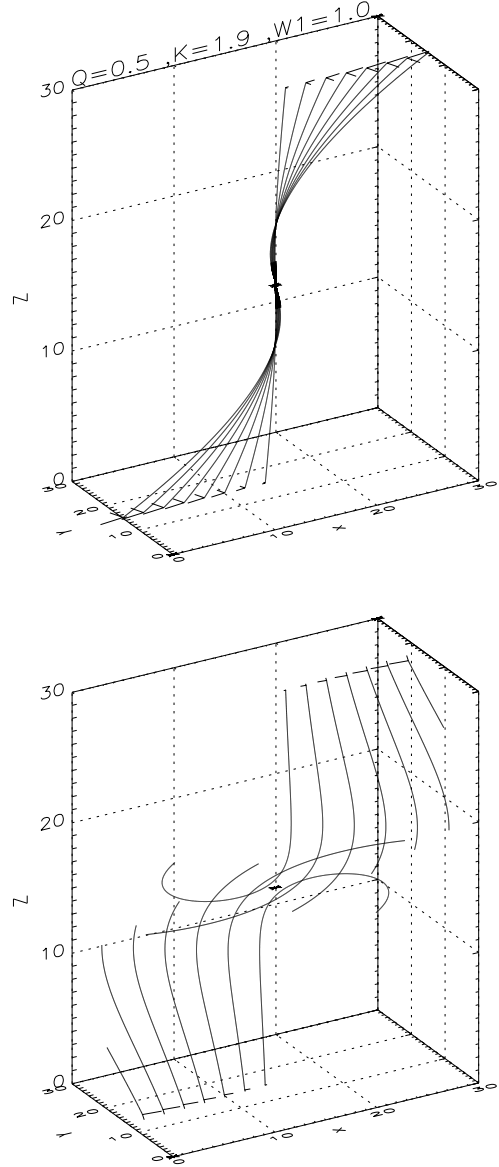,width=8cm}}
\caption{Stream lines (upper panel) and magnetic field lines (lower
panel)  of the fields
corresponding to Eqs.~\ref{sol_rot_mag1} ($B=0$)
 with parameters corresponding to
 the weak rotation regime (first case). The converging center ($r=0$) is located in the
 box centre ($x=15,y=15,z=15$). The gas is accreted at the pole and at 
the equator.}
\label{lig_champ1}
\end{figure}

\begin{figure}[htbp]
\centerline{\psfig{file=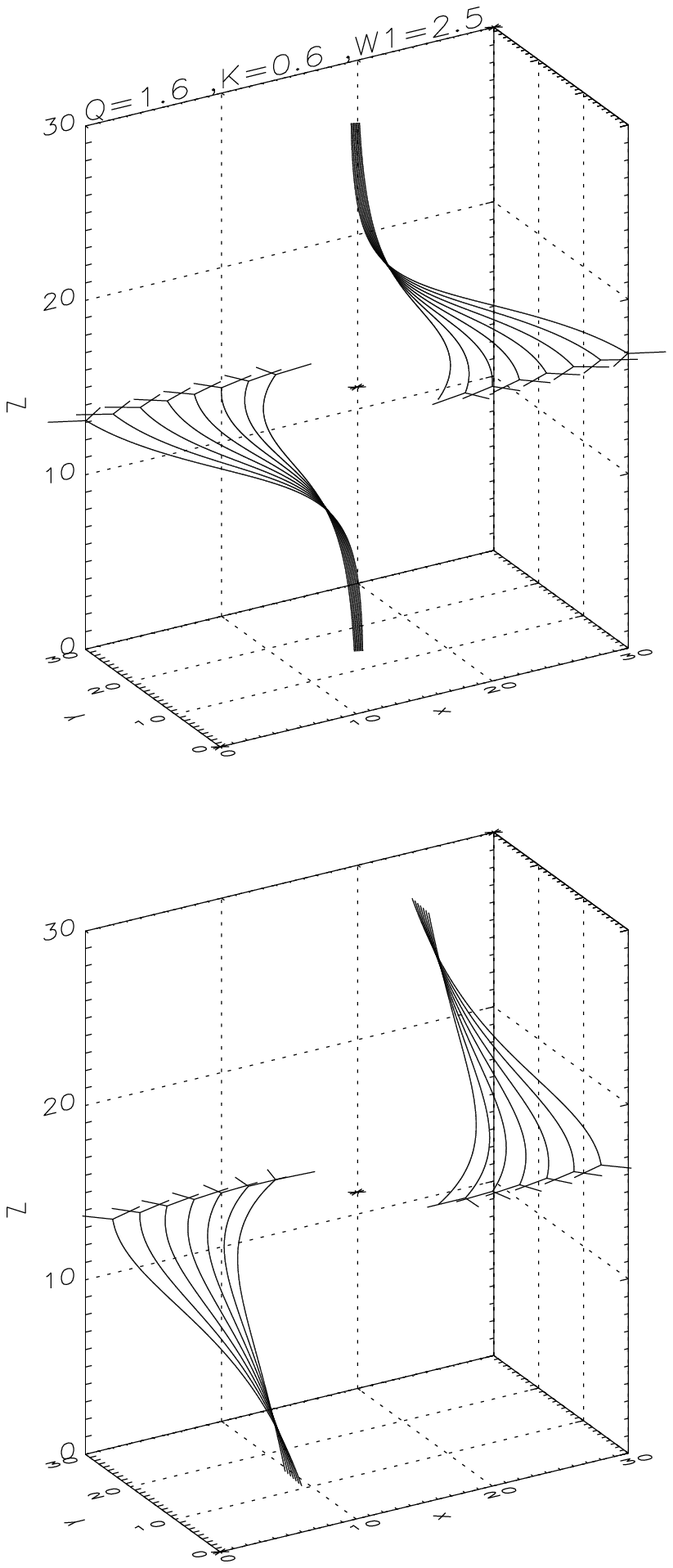,width=8cm}}
\caption{Same as Fig.~\ref{lig_champ1} for parameters corresponding to the 
second case (strong rotation): $2 QK < W_1 < 3QK$. The gas is accreted at the equator and ejected  at the pole.}
\label{lig_champ2}
\end{figure}

In this section we consider a rotating and magnetized cloud. 
We look for solutions stated by 
Eqs.~\ref{hypo},~\ref{mag_sol} and:
\begin{eqnarray}
h_2 (\theta) = Q \sin \theta.
\label{mag2_sol}
\end{eqnarray}
The corresponding azimuthal magnetic component is proportional to:
\begin{eqnarray}
B _\phi \propto \sin \theta \; r. 
\label{toroid}
\end{eqnarray}

Eqs.~\ref{cmat_red},~\ref{cmom1_red},~\ref{cmom2_red} and~\ref{cmom3_red}
 lead to:
\begin{eqnarray}
\left\{
\begin{array} {l}
  U_1 =  3 V_ 0  -2 \omega, \\
V _0 ^2  - W_1 ^2  - \omega V_0 =
   - 1 / 3 - 2 \phi _1
  - 2 Q ^2, \\
- 2 V_0 U _1 + U _1 ^2 + W_1 ^2 + 
\omega U _1 =    6 \phi _1 + 2 Q ^2, \\
V _0 W _1  -  \omega W _1 / 2  = K Q U _1 ,
\end{array}
\right.
\label{cond_mom_rot_mag}
\end{eqnarray}

whereas Eq.~\ref{m2_red} gives the  same constraint
 as Eq.~\ref{cmat_red}.

\begin{figure}[htbp]
\centerline{\psfig{file=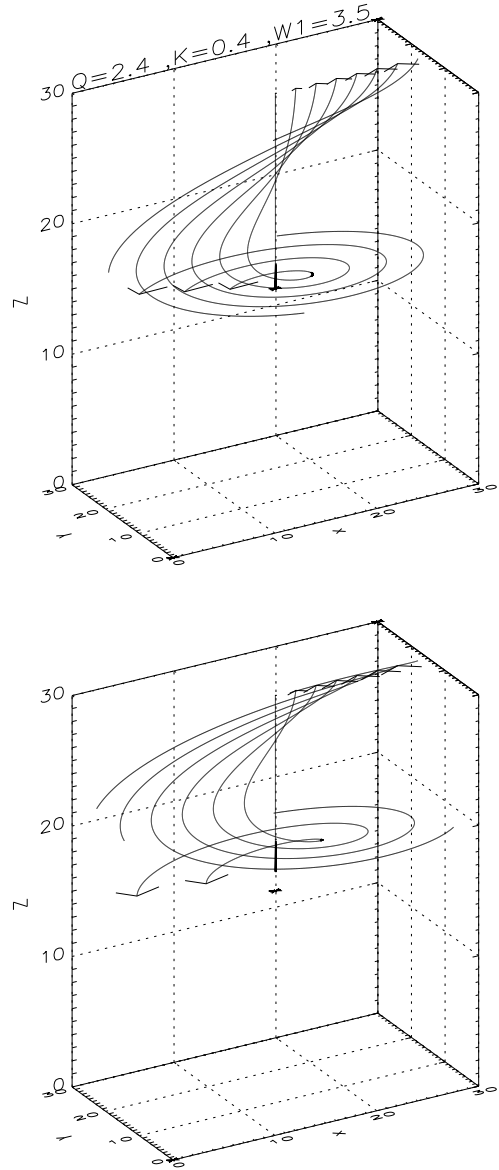,width=8cm}}
\caption{Same as Fig.~\ref{lig_champ1} for parameters corresponding to the 
third case (strong rotation): $3 QK < W_1 < 4QK$.
 The gas is accreted at the pole and ejected at the equator.}
\label{lig_champ3}
\end{figure}

\begin{figure}[htbp]
\centerline{\psfig{file=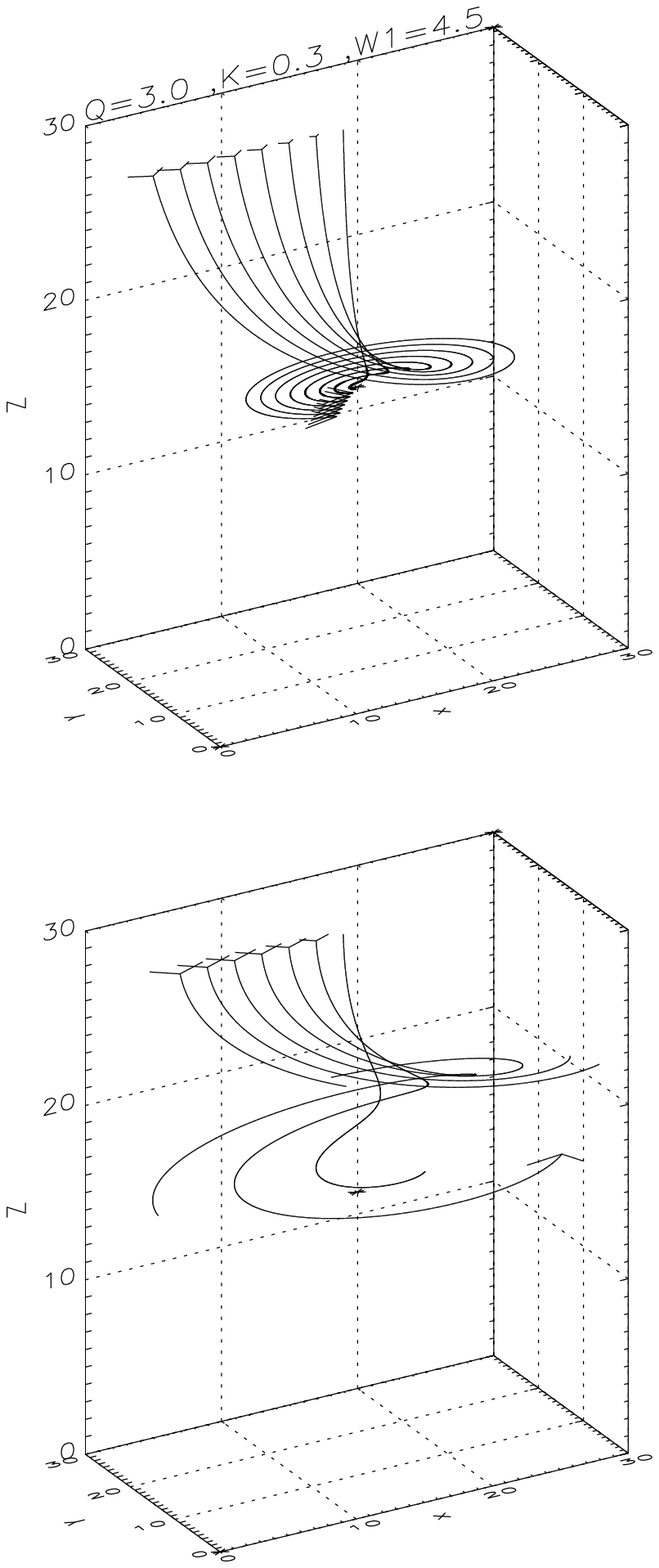,width=8cm}}
\caption{Same as Fig.~\ref{lig_champ1} for parameters corresponding to the 
fourth case (strong rotation): $4 QK < W_1 < 6 QK$. 
The gas is accreted at the equator and  at the pole.}
\label{lig_champ4}
\end{figure}

The fourth equation derive from the azimuthal momentum conservation,
 its  right-hand side is the  magnetic tension which is proportional
to the product of the poloidal and toroidal field.
 The term $Q ^2$ in the right-hand side of the second and the third
equations is the Lorentz force due to the toroidal field. 
It tends to compress the gas
because the magnetic toroidal intensity increases with the distance to the
z-axis, $\varpi$. 
There is no magnetic term due to the poloidal field only 
(proportional to $K^2$)
 because its contribution  vanishes (see Eq.~\ref{m1_nul}).

As in the non rotating case, one can easily check that adding a uniform
 magnetic field (Eq.~\ref{champ_cons}) is still possible (the terms 
$(-\sin \theta) \partial _\theta (\sin \theta B _\phi) + \cos \theta \sin \theta \partial _r
(r B _\phi)$ in Eq.~\ref{t_consmom3} and 
$\partial _r(r V _\phi \cos \theta) + \partial _\theta ( V _\phi (-\sin \theta) )$
in Eq.~\ref{eqmag} cancel). 
The physical meaning is that a solid body rotation does not stretch a uniform
field parallel to the rotation axis and an helicoidal field (uniform poloidal
field along the z-axis and toroidal field proportional to $\varpi$) does not
induce any  azimuthal tension. 

Consequently, one obtains the poloidal magnetic field given by 
Eqs.~\ref{magrad}-\ref{magort} whereas the toroidal field is given by Eq.~\ref{toroid}.

The solution of Eq.~\ref{cond_mom_rot_mag} is given by:
\begin{eqnarray}
\left\{
\begin{array} {l}
\omega  = \pm \sqrt{ { (2 + 8 Q ^2 - 4 W _1 ^2) 
 (W _1 - 3 Q K )^2 \over 
(W _1 - 2 Q K) (W _1 - 6 Q K) } } \; , \; \\
\phi _1 =  1 / 12  + \omega ^2 
{  Q K W _1  -2 (Q K )^2 \over 4  (W _1 - 3 Q K )^2 } \; , \;
 U _1 = -  { W _1 \over   W _1 - 3 Q K } \omega / 2
\; , \\
V(\pi /2) = V _0 =  { W _1 / 2
  -2 Q K \over  W _1 - 3 Q K  }  \omega
\; , \; 
V(0) =   { W _1 
  -2 Q K \over  W _1 - 3 Q K  }  \omega.
\end{array}
\right. 
\label{sol_rot_mag1}
\end{eqnarray}

$W _1$, $Q$ and $K$ are free parameters.
The transformation $W _1 \rightarrow - W _1 \; , \; Q  K \rightarrow -
Q K $ leaves the solution invariant and we can assume without restriction
 that $W _1 > 0$. \\

Two cases arise (conditions for $\omega ^2$ to be positive): \\

- $W _1 <\sqrt{ 1 /2 + 2 Q ^2} $ (weak rotation) and 
$W _1 < 2 Q K $ or $ W _1 > 6 Q K$.
The centrifugal force is smaller than the contribution of the gravity and
the toroidal magnetic forces. 

The gravitational potential is spherical ($\phi _1=0$)  if 
$W _1 + 12 QK (2 Q ^2 - W _1 ^2)=0$. 
We have $ V(0) \omega > 0$ and $V(\pi /2) \omega > 0$ (inflow). 

 If $Q K U _1 \le 0$, the azimuthal magnetic 
tension tends to decrease the azimuthal velocity 
(see Eqs.~\ref{cond_mom_rot_mag}). This is the magnetic braking effect.
If $Q K U _1 \ge 0$, the magnetic tension increases the rotation velocity. 
This forcing could be related to the Torsional Alfv\'en waves inducing
 collapse studied by Habe et al. (1991). 

This regime is illustrated in
Fig.~\ref{lig_champ1} that displays the stream lines and magnetic field lines 
for $W_1=1, Q=0.52$ and $K=1.9$. \\

- $W _1 > \sqrt{ 1 /2 + 2 Q ^2 }$ (fast rotation) and 
$ 2 Q K <  W _1 < 6 Q K$. 
The centrifugal force dominates gravity (more precisely the spherical part 
of the gravitational force) and magnetic toroidal pinching.
$\phi _1$ is greater than $1/12$ and thus gravity tends to move away
 from the origin the gas at the equator (see sect.~\ref{gravi}).

Assuming that, $ 2 QK > \sqrt{1/2 + 2 Q ^2}$,   three cases have to be distinguished:
\begin{description}
\item[-] $ 2 QK < W _1 < 3 QK $ then 
$ V(0) \omega < 0$ and $V(\pi /2) \omega > 0$, \\
 $\omega < 0$ (collapse) implies that the gas is ejected at the pole
(outflow) and accreted at the equator (inflow).
Fig.~\ref{lig_champ2} displays the stream lines and magnetic field lines 
for $W_1=2.5, Q=1.6$ and $K=0.6$.

As for the weak rotation case  with $Q K U _1 \le 0$, 
the azimuthal magnetic tension tends to
 brake the clouds.

Since $\phi _1 > 1 /12$, the gravitational potential becomes positive at 
the pole and gravity is responsible of the gas ejection along the pole
(gravity is the only force along the z-axis).
It is not clear if this is the  same physical mechanism that is responsible
 of  outflows 
 found in numerical simulations of collapse-driven outflow (e.g. Tomisaka 1998, 2001).

\item[-]  $ 3 QK < W _1 < 4 QK $ then 
$ V(0) \omega > 0$ and $V(\pi /2) \omega < 0$, \\
 $\omega < 0$ (collapse) implies that the gas is accreted (inflow) at the pole
and ejected (outflow) at the equator.
Fig.~\ref{lig_champ3} displays the stream lines and magnetic field lines 
for $W_1=3.5, Q=2.4$ and $K=0.4$,

Since $U _1 Q K \ge 0$, the magnetic tension tends to increase the cloud 
rotation velocity. 
 
\item[-]  $ 4 QK < W _1 < 6 QK $ then 
$ V(0) \omega > 0$ and $V(\pi /2) \omega > 0$,
 $\omega < 0$ (collapse) implies that the gas is ejected at the pole
and at the equator (usual behaviour).
Fig.~\ref{lig_champ4} displays the stream lines and magnetic field lines 
for $W_1=4.5, Q=3$ and $K=0.3$.

As for the previous case, the magnetic tension tends to increase the cloud
rotation velocity.

\end{description}

In the limit $W _1 \rightarrow 3 QK$, one finds the surprising result
that $\omega \rightarrow 0$ (stationary solution) but, if 
$18 Q ^2 K ^2 - 4 Q ^2 -1 > 0$:
\begin{eqnarray}
\begin{array} {l}
|V (\pi/2)| \rightarrow  1 / \sqrt{6} \sqrt{ 18 Q ^2 K ^2 - 4 Q^2 -1}    \ne 0 , \\
 |V (0)| \rightarrow   \sqrt{2/3} \sqrt{ 18 Q ^2 K ^2 - 4 Q^2 -1}
\ne 0, \\
 |U _1 | \rightarrow   \sqrt{3/2} \sqrt{ 18 Q ^2 K ^2 - 4 Q^2 -1}
\ne 0. 
\end{array} 
\label{sol_statio}
\end{eqnarray}
with $V(0) V(\pi /2) < 0$ and where 
the sign of $V(0)$ can be positive or negative. 
The physical meaning is that the outflow rate is equal to the inflow rate.
 Indeed the accretion rate, $J$, is proportional to:
\begin{eqnarray}
J (r) = 2 \int _0 ^ {\pi /2}  2 \pi r^2 \sin \theta \rho V_r d \theta \propto 
V_0 - U _1 / 3
\end{eqnarray}
  and consequently vanishes (see Eqs.~\ref{sol_statio}).

Let us summarize the results.
\begin{description}
\item [-] We find four different regimes, one corresponds to a low 
rotation and three to a strong rotation 
(greater than gravity and magnetic toroidal forces). 
For two sets of parameters, one finds outflows
at the pole or at the equator. 
\item[-] In the weak rotation regime with $Q K U _1 \le 0$ and in the first case of
the strong rotation regime, the magnetic tension brakes the cloud. In the other
cases, it tends to accelerate the cloud rotation. 
\end{description}

Although the magnetic field stated by Eqs.~\ref{magrad},~\ref{magort} 
and~\ref{toroid} is not force-free, since the azimuthal tension and
the magnetic toroidal pinching do not vanish, the poloidal magnetic
pressure is equal to zero and does not support the cloud as it is the case
in the numerical simulations of gravo-magnetic collapse which generally assume
uniform initial mass-to-magnetic flux ratio. 
This fact clearly limits  the applicability of the 
solutions to collapse for which the magnetic poloidal support is not 
dominant (weak magnetic field or field dominated by toroidal component).

\section{Aspherical temperature}

In this section we briefly consider the case of an aspherical thermal pressure.
We also extend the angular dependence of the various fields
and find   solutions that present velocity fields  not symmetrical with
respect to the equatorial plan.  

With the uniform density assumption, as we already said in
 Sect.~\ref{pression thermique}, 
this means that the temperature 
is not a function of the density only as it is usually assumed.
Although less astrophysically relevant than the uniform temperature
 cases studied in the previous section, such 
temperature gradients are expected when radiative processes
 are considered.

\subsection{Results}

We look for solutions given by:
\begin{eqnarray}
\left\{
\begin{array} {l}
V(\theta) = V_0 - U _1 \cos ^2 \theta - U_2 \cos \theta \sin \theta \;
, \;  \\
U(\theta) =    U _1 \cos \theta \sin \theta  + U _2 \sin ^2 \theta \; , \; \\
\Pi (\theta) = \Pi _0 + \Pi _1 \cos ^2 \theta + \Pi _2 \cos \theta \sin \theta
 \; , \;  \\
h _1 (\theta) = K( U _1 \, \cos \theta \sin ^2 \theta +
U _2 \, \sin ^3 \theta) \; , \; \\
W (\theta) = W _1 \sin \theta \; , \; h _2 (\theta) = Q \sin \theta.
\end{array}
\right.
\label{hypo_ad}
\end{eqnarray}

We then have:
\begin{eqnarray}
\begin{array}{l}
B _r \propto ( - 3 U _1 \cos ^ 2 \theta + U _1  - 
3 U _2 \cos \theta \sin \theta) \; r,   \\
B _\theta \propto (3 U _1 \cos \theta \sin \theta + 3 U _2 \sin ^2 \theta)\; r, \\
B _ \phi \propto \sin \theta \; r.
\end{array}
\end{eqnarray}
As  physically expected, the azimuthal and orthoradial components 
of the magnetic and velocity fields vanish at the pole.

Eqs.~\ref{cmat_red},~\ref{cmom1_red},~\ref{cmom2_red} and~\ref{cmom3_red} give:
\begin{eqnarray}
\left\{
\begin{array} {l}
  U_1 =  3 V_ 0  -2 \omega, \\
V _0 ^2 - \omega V _0 - W _1 ^2
= \\ 
\hspace{1.5cm} - 2 \Pi _0 - 1 / 3 - 2 \phi _1   -9 K^2
U_2^2 - 2 Q ^2 \; , \; \\
- 2 V _0 U _1 + U_1 ^2 + \omega U_1 + W _1 ^2 = \\
 \hspace{1.5cm} - 2 \Pi _1  + 6
\phi _1 + 9 K^2 U _2 ^2 + 2 Q ^2
  \; , \; \\
-2 V_0 U _2 + U_2 U _1 + \omega U _2 =- 2
\Pi _2 -9 K^2 U _1 U_2  \; , \; \\
\Pi _2 + 6 K ^2 U _1 U_2 = 0 \; , \; \\
V _0 W _1 - \omega W _1 / 2  = K Q U _1.
\end{array}
\right.
\label{cond_mom_ad}
\end{eqnarray}
The second, third and fourth equations derive from the radial momentum
conservation and  the third, the fourth and the fifth  from the
orthoradial momentum conservation. The fifth equation shows that there
is a non vanishing orthoradial component of the Lorentz force at the pole 
 that must be counterbalanced by the thermal pressure. 

Eq.~\ref{cen_red} leads to:

\begin{eqnarray}
\left\{
\begin{array} {l}
(-2 U_1) \Pi _0 + ( (2 \gamma -4) \omega + 2 V _0 - 2 U _1 ) \Pi _1 + \\
 \hspace{5cm} U _2 \Pi _2 =0 \; ,\\
(- 2 U_2) \Pi _0 + (-2 U_2) \Pi _1 + \\
 \hspace{2cm} ( (2 \gamma -4) \omega + 2 V_0 - U _1 ) \Pi _2 =0 \; ,\\ 
 ((2 \gamma -4) \omega + 2 V _0) \Pi _0 + (-U _2) \Pi _2 =0 .
\label{eq_press_ad}
\end{array}
\right.
\end{eqnarray}

One obtains  the four solutions:
\begin{eqnarray}
\left\{
\begin{array} {l}
\Pi _0 = - \Pi _1\; , \; (\gamma -2) \omega + V _0 = 0  \; , \;
\Pi _2 =0, U _2 =0 \; , \\
\Pi _0 = 0 \; , \; (\gamma -2) \omega + V _0 - U _1 = 0 \; , \;
 \Pi _2 =0, U _2 =0 \; , \\
 U_2 \Pi _2 = 2 U _1 \Pi _0 \; , \;  
(U _1 ^2 - U _2 ^2 ) \Pi _2 = 2 U _1 U _2 \Pi _1 \; , \; \\
 \hspace{4cm} (\gamma -2) \omega + V _0 = U_1   \; ,  \\ 
 \Pi _0 + \Pi _1 =0 \; , \; U_2 \Pi _2 =  U _1 \Pi _0 \; , \;
 2(\gamma -2) \omega + 2 V_0 = U _1. 
\label{press_ad}
\end{array}
\right.
\end{eqnarray}

The  first and the second solutions  are symmetrical with respect
 to the equatorial plane whereas the third and the fourth  are not.

\subsection{Non-symmetrical solutions with respect to the equatorial plan}
In this section we give and discuss the two solutions of
Eqs.~\ref{cond_mom_ad}  and Eqs.~\ref{press_ad}  
with $U_2 \ne 0$ and $\Pi _2 \ne 0$ as they present interesting 
physical features.  
The solutions symmetrical with respect to the equatorial plan are given 
for completeness in appendix B.

These two solutions are given by:

\begin{eqnarray}
\left\{
\begin{array} {l}
\Pi _0 + \Pi _1 = - 3 K ^2 U _1 ^2 \; ,  
 U_2 \Pi _2 = 2 U _1 \Pi _0 \; , \;  
\Pi _0 = {2 - \gamma \over 3 \gamma -4 } U _2 ^ 2  \; , \\
\omega = \pm \sqrt{ { 1 - 2 \Pi _0  + 4 Q ^2 - 2 W _1 ^2 \over 2 - \gamma 
}} \; , \; \\
\phi _1 =  1 /12 -  \omega ^2  ( 2 - \gamma ) ^2 / 8 
   \; , \\
 K ^2 = {1 \over 3} {2 - \gamma  \over 4 - 3
\gamma } \; , \; W _1 =  K Q { 3 \gamma -4 \over \gamma -1 }
\; , \;  U_1 = (3 / 2 \gamma -2) \omega  \; , \\
V(\pi /2) = V_0 =  \gamma  \omega / 2   \;  , \; V(0) =  (2 - \gamma) \omega.
\end{array}
\right. 
\label{sol_ad1}
\end{eqnarray}

\begin{eqnarray}
\left\{
\begin{array} {l}
\Pi _0 + \Pi _1 =0
\; , \; U_2 \Pi _2 =  U _1 \Pi _0 \; , \;  \Pi _0 = - 6 K ^2 U _2 ^2 \; , \\
\omega = \pm \sqrt{ -1  - \Pi _0  -4 Q^2 + 2 W _1 ^2 \over 2 (2 \gamma -3) (6 \gamma -7)}
 \; , \; \\
 \phi _1 =  1 / 12 +  \omega ^2 (2 \gamma  - 3)
(4 \gamma - 5 ) / 2  \; , \\
 \; K ^2 = { 1 \over 6 } {3 - 2 \gamma  \over 4 - 3 \gamma  } \; , \;
W _1 = 4 K Q { 4 - 3 \gamma \over 5 - 4 \gamma } \; , \; 
U_1 = 2 (3  \gamma -4) \omega  \; , \; \\
V(\pi/2) = V_0 = 2(\gamma -1) \omega \; , \;  V(0) = 2 (3 - 2 \gamma) \omega.
\end{array}
\right.
\label{sol_ad2}
\end{eqnarray}

For the first solution (Eqs.~\ref{sol_ad1}), $K ^2 > 0 \Rightarrow$ $\gamma < 4 /3$, 
$\phi _1 < 1 /12$ and $W _1 ^2 < 1/2 - \Pi _0 + 2 Q ^2$. 

The rotation parameter, $W _1$, is a function of the toroidal magnetic
intensity, $Q$, and the polytropic index $\gamma$.
The rotation tends to prevent the condensation   whereas the magnetic field,
tends to compress the gas
(see the terms $2 W _1^2$ and $-4 Q  ^2$ in the expression of $\omega$)
 and it is worthwhile to know the global trend of both effects.  
 We have:
\begin{eqnarray}
-4 Q ^2 + 2 W _1 ^2 = -{ 2 / 3} \; Q ^2 { 3 \gamma ^2 - 2\gamma - 2
\over (\gamma -1) ^2 }.
\end{eqnarray} 
 If $\gamma < (1 + \sqrt{7}) / 3$ one has~: $-4 Q ^2 + 2 W _1 ^2 > 0$
 which means that  
for the solution stated by Eqs.~\ref{sol_ad1}, the global 
effect of the rotation and the magnetic compression is to prevent
 the collapse.
 If $\gamma > (1 + \sqrt{7}) / 3$ one has~: $-4 Q ^2 + 2 W _1 ^2 < 0$.  \\

For the second solution (Eqs.~\ref{sol_ad2}), one has: 
 $K ^2 > 0 \Rightarrow $ $\gamma < 4 /3$ or $\gamma > 3/2$.

 We have:
\begin{eqnarray}
- 4 Q ^2 + 2 W _1 ^2 = - 4 /3 \;  Q ^ 2 
{ 24 \gamma ^2 - 52 \gamma + 27 \over (5 - 4 \gamma ) ^ 2 } 
\label{signe_cond1}
\end{eqnarray}
and  
\begin{eqnarray}
- 4 Q ^2 + 2 W _1 ^ 2 > 0 \Leftrightarrow (13 - \sqrt{7}) / 12 < \gamma
<  (13 + \sqrt{7}) / 12 .
\label{signe_cond2}
\end{eqnarray}

Thus if $(13 - \sqrt{7}) / 12 < \gamma <  (13 + \sqrt{7}) / 12 $ then
$-4 Q ^2 +  2 W _1 ^2  -  \Pi _0 -1$ can be positive even with $\Pi _0 >0$
if $Q$ is large enough, whereas in the
opposite case, only negative values of $\Pi _0$ 
(the thermal pressure tends to compress the gas)
 are possible in order for
$-4 Q ^2 +  2 W _1 ^2  -  \Pi _0 -1$ to be positive. 

Two cases arise: \\

-  $\; -4 Q ^2 +  2 W _1 ^2  -  \Pi _0  < 1$ (weak magnetic intensity/rotation and/or weak
aspherical thermal pressure), 
 implies $7/6 < \gamma < 4/3$. 

One has $V(0) \omega > 0$ and 
$V(\pi /2) \omega > 0$. \\

-  $\; -4 Q ^2 +  2 W _1 ^2 -  \Pi _0  > 1 $ (strong magnetic
intensity/rotation  and/or strong aspherical thermal pressure), 
 implies $\gamma < 7 / 6$ or  $\gamma > 3/2$.

\begin{description}
\item[-] If $\gamma < 1$, one has  $V(\pi /2) \omega < 0$ and  $V(0) \omega >
0$,  
 $\omega < 0$ (collapse) implies that the gas is accreted (inflow) at the pole
and ejected (outflow) at the equator.
\item[-] If $1< \gamma < 7/6$, $\omega <0$ (collapse) implies that the gas 
is accreted at the pole and at the equator (inflow).
\item[-] If $\gamma > 3 /2$,  $V(\pi/2) \omega > 0$ and  $V(0) \omega < 0$, 
$\omega < 0$ (collapse) implies that the gas is ejected (outflow) at the pole
and accreted (inflow) at the equator.
\end{description}

As for Sect.~\ref{rot_mag}, one finds that the outflow solutions have $\phi _1 > 1/12$ (positive gravitational potential at the pole).



\section{Conclusion}
In this paper, we apply the transformation proposed by Munier \& Feix
(1983) to the axisymmetric magnetohydrodynamical equations 
   of  a self-gravitating polytropic gas.
We  reduce these equations to a system of 8 
ordinary differential equations of the variable $\theta$
(with Eq.~\ref{mag_sol} it is reduced to 7 equations only).
  We then  derive exact and explicit self-similar solutions of these 
 ordinary differential equations capable to
describe the gravitational collapse of a uniform density cloud  with 
aspherical pressure, aspherical gravitational potential,  magnetic
field and rotation. These solutions generalise the Larson-Penston solutions
 in the case where these processes are considered and thus describe the
subsonic core of a collapsing cloud (Whitworth \& Summers 1985).

For the solutions of the collapse with an aspherical gravitational 
potential, we show that the sonic surface is an ellipsoid that can be oblate
or prolate. Assuming that a bifurcation
occurs at this surface (as it is the case for the spherical solution),
 this suggests that these solutions describe the internal part
of elliptical clouds.

We find that the solutions  given by
Eqs.~\ref{bouquet_sol},~\ref{sol_hyd1},~\ref{sol_hyd2}
and~\ref{sol_hyd3}  describing the collapse of a non rotating cloud,
 are compatible with a  magnetic field given by
Eqs.~\ref{magrad} and~\ref{magort} 
 of any intensity since the 
magnetic configuration is such that the Lorentz force vanishes.

The solutions describing the collapse of  magnetized and rotating
clouds with uniform thermal pressure (Eqs.~\ref{sol_rot_mag1}) 
present four regimes, one corrresponding to slow rotation and three to
rapid rotation. The solutions corresponding to these three regimes 
(rapid rotation) present
(depending on the rotation and on the magnetic intensity)
 outflow at the pole, outflow  at the equator or inflow only.
 There is a value of the rotation (Eqs.~\ref{sol_statio})
for which the outflow rate is equal to the inflow rate and the
solution is stationary ($\omega=0$).  

Four regimes, depending on $\gamma$, are also found for one of the (non symmetrical 
with respect to the equatorial plane) solutions 
describing the collapse of a magnetized and rotating cloud with aspherical pressure
(Eq.~\ref{sol_ad2}). One regime ($\gamma <1$) presents outflow at the equator,
one regime ($\gamma > 3/2$) outflow at the pole.

The two main restrictions of the present work are, first
  the Lorentz force due to the poloidal magnetic field  vanishes (if $U _2 =0$)
 and thus, the magnetic poloidal force
 is not  a support against the gravitational
collapse (if $U _2 \ne 0$, it compresses the gas),
 second the solutions diverge when $r$ goes to infinity and 
  are valid in a finite domain only. 
The study of the bifurcation that could occur at the sonic 
(or magneto-sonic) point has not 
been considered here but must be carefully addressed.
Indeed, a numerical solution of the two dimensional stationary system
of Eqs.~\ref{t_consmat}-\ref{t_eqmaga2} (see e.g. Arthur \& Falle 1991 who
carry out a similar study for a bidimensional self-similar supernovae
 explosion) must be addressed at some stage. 
 It should make possible to avoid the vanishing poloidal force and 
 should give the solution on the whole domain.
Such a study should give a sensible hint of the gravo-magnetic condensation.

Nevertheless, 
these  solutions that generalise the Larson-Penston solution 
 have an explicit dependence on the radius $r$
and the colatitude $\theta$ and present a wide class of physical behaviours.
They  offer an opportunity for the testing of numerical codes and 
a  starting point for future analytical studies. \\

{\it I thank Serge Bouquet for helpful and stimulating discussions.
I thank Frank Shu for his help in improving the manuscript and Shantanu Basu
 for interesting suggestions and constructive remarks.
 I also acknowledge a critical and helpful
 reading of the manuscript by Serge Bouquet, Anthony Whitworth, 
 Robin Williams and Michel P\'erault.}

\appendix

\section{Exact and explicit solutions of the gravo-magnetic condensation}
In this appendix we summarize the solutions found in this paper and
give them  explicitly. The values of the parameters $\gamma$, $\omega$, $\Pi
_0$, $\Pi _1$, $\Pi _2$, $U _1$, $U _2$,  $V _0$, $\phi _1$, $K$
and $Q$ being given by
Eqs.~\ref{bouquet_sol},~\ref{sol_hyd1},
~\ref{sol_hyd2},~\ref{sol_hyd3},~\ref{sol_rot1},~\ref{sol_rot2},~\ref{sol_rot3},
\ref{sol_rot_mag1},~\ref{sol_rot_mag2},~\ref{sol_rot_mag3},~\ref{sol_ad1}
or~\ref{sol_ad2},  $\rho _0$ and $P _0$ being free parameters and
$\Omega _0 = \sqrt{ 4 \pi G \rho _ 0}$, 
 Eqs.~\ref{consmat}-\ref{eqmag} admit the solutions given by:
\begin{eqnarray}
\rho ( t,r, \theta)= { \rho _0 \over (1 + \Omega _0 \omega t) ^2 }, 
\label{s1}
\end{eqnarray}

\begin{eqnarray}
\Phi ( t,r, \theta)=  { \Omega _0 ^2 \over (1 + \Omega _0 \omega t) ^2}
 ( - {1 \over 6} + 3 \phi _1 \cos ^2 \theta - \phi _1 ) \; r ^2,
\label{s2}
\end{eqnarray}

\begin{eqnarray}
\nonumber
 P ( t,r, \theta) &=& { P _0 \over   (1 + \Omega _0 \omega t) ^ {2 
\gamma } } + 
 { \rho _0 \Omega _0 ^2 \over (1 + \Omega _0 \omega t) ^4 } \times \\
 &&   (\Pi _0 + \Pi _1 \cos ^2 \theta +
\Pi _2 \cos \theta \sin \theta ) \;  r ^2,
\label{s3}
 \end{eqnarray}

\begin{eqnarray}
\nonumber
V _r ( t,r, \theta) &=&  \\
 { \Omega _0 \over 1 + \Omega _0 \omega t } &\times&
(V _0 - U _1 \cos^2 \theta - U _2 \cos \theta \sin \theta ) \;  r,  
\label{s4}
\end{eqnarray}

\begin{eqnarray}
V _\theta( t,r, \theta)={ \Omega _0 \over 1 + \Omega _0 \omega t }
( U _1 \cos \theta \sin \theta + U _2  \sin ^2 \theta ) \; r,
\label{s5}
\end{eqnarray}

\begin{eqnarray}
V _\phi( t,r, \theta)={ \Omega _0 \over 1 + \Omega _0 \omega t }
W _1 \sin \theta \; r, 
\label{s6}
\end{eqnarray}

\begin{eqnarray}
\nonumber
B _r ( t,r, \theta) &=& {\sqrt{\mu _0 \rho _0} \Omega _0 \over (1 + \Omega _0
\omega t) ^2 } \times \\
& K& ( - 3 U _1 \cos ^2 \theta + U _1 - 3 U _2 \cos \theta
\sin \theta ) \;  r,
\label{s7}
 \end{eqnarray}

\begin{eqnarray}
\nonumber
B _\theta ( t,r, \theta) & =&  
 {\sqrt{\mu _0 \rho _0} \Omega _0 \over (1 + \Omega _0
\omega t) ^2 } \times \\
 &K& (  3 U _1 \cos  \theta \sin \theta +  3 U _2 \sin ^ 2
\theta ) \;  r,
\label{s8}
\end{eqnarray}

\begin{eqnarray}
B _\phi ( t,r, \theta) = 
  {\sqrt{\mu _0 \rho _0} \Omega _0 \over (1 + \Omega _0
\omega t) ^2 } 3 Q  \sin \theta  \;  r.
\label{s9}
\end{eqnarray} \\

With $U _2 = \Pi _2 = 0 $ ($\gamma, \omega, \Pi _0, \Pi _1, U _1,
 V _0, W _1, Q$  and
$\phi _1$ being given by Eqs.~\ref{bouquet_sol},~\ref{sol_hyd1},
~\ref{sol_hyd2},~\ref{sol_hyd3},~\ref{sol_rot1},~\ref{sol_rot2},~\ref{sol_rot3},
~\ref{sol_rot_mag1},~\ref{sol_rot_mag2},~\ref{sol_rot_mag3}),
 B being a free parameter, Eqs.~\ref{consmat}-\ref{eqmag} 
admit the solutions given by Eqs.~\ref{s1}-\ref{s6} and~\ref{s9} and:

\begin{eqnarray}
\nonumber
B _r ( t,r, \theta) &=& 
 { B \over (1 + \Omega _0 \omega t ) ^{2 V _0 / \omega} }  \cos \theta + \\
&& { \sqrt{\mu _0 \rho _0} \Omega _0 \over  (1 + \Omega _0
\omega t) ^ {2}}   K ( - 3 U _1 \cos ^2 \theta + U _1  ) \; r \; ,
\end{eqnarray}
\begin{eqnarray}
\nonumber
B _\theta ( t,r, \theta) &=& 
-  { B \over (1 + \Omega _0 \omega t ) ^{2 V _0 / \omega} }  \sin \theta + \\
& & { \sqrt{\mu _0 \rho _0} \Omega _0 \over  (1 + \Omega _0
\omega t) ^{2 }} K (  3 U _1 \cos  \theta \sin \theta  ) \; r.
\end{eqnarray}

\section{Aspherical temperature field}
In this appendix we give for completeness the solutions of 
Eqs.~\ref{cond_mom_ad} and~\ref{press_ad} that are
symmetrical with respect to the equatorial plan ($U_2=0, \Pi_2=0$)
 and briefly discuss their main  properties.

\subsection{Aspherical pressure}
Two  solutions with $W _1=K=Q=0$    are obtained:
\begin{eqnarray}
 \left\{
\begin{array} {l}
\Pi _0 = - \Pi _1  \;  , \;  
 \Pi _0 = 
 {3 \phi _1 \gamma - 5 \phi _1 - 1 / 3  + \gamma / 4
\over 3 - 2 \gamma } \; , \; \\
\omega = \pm \sqrt{  2 \phi _1 
 - 1/6  \over
  (1 - \gamma ) (3 - 2 \gamma) } \; , \; 
 U _1 = (4 - 3 \gamma) \omega \; , \\
V(\pi /2) =  V_0 = (2-\gamma) \omega \; , \; 
V(0) = 2(\gamma -1) \omega.  
\end{array}
\right. 
\label{sol_hyd2}
\end{eqnarray}

\begin{eqnarray}
 \left\{
\begin{array} {l}
\Pi _0 = 0 \;  , \;  
\Pi _1 = - 1 / 2 +    (3 \gamma -2) 
(2 - \gamma) \omega ^2 /4 \; , \; \\
\omega = \pm \sqrt{  4 / 3 + 8 \phi _1   \over 
 \gamma (2 - \gamma ) } \; , \; 
 U _1 =  (3 \gamma -4) \omega / 2 \; , \\
V (\pi/2) = V _0 =  \gamma  \omega / 2 \; , \;
V( 0) = (2 - \gamma ) \omega.
\end{array}
\right.
\label{sol_hyd3}
\end{eqnarray}
These two solutions describe a collapse or an expansion with an
  aspherical thermal pressure   
(Fig.~\ref{champ_sym} displays the velocity field of the solution stated
by Eqs.~\ref{hypo}-\ref{sol_hyd2} for $\gamma=5/3$, $\gamma=4/3$ and
 $\gamma = 1.2$). 
The gravitational 
potential can be uniform ($\phi _1=0$) or aspherical ($\phi _1 \ne
0$). In the limit $\phi _1 \rightarrow 0$ and $\gamma \rightarrow
4/3$, one recovers the solution stated by Eq.~\ref{bouquet_sol}.

For the first solution (Eqs.~\ref{sol_hyd2}),
 a solution exists only if $\phi _1 < 1/12$ and $1 < \gamma < 3/2$
 or if $\phi _1 > 1/12$ and
$\gamma <1$ or $\gamma > 3/2$. 
If $\gamma < 1$ and $\phi _1 > 1/12$, the solution presents  outflows
at the pole. 

For the second solution (Eqs.~\ref{sol_hyd3}), one has $\phi_1 > -1 /
6$  and $0<\gamma<2$.

\subsection{Aspherical pressure and rotation}
Two  solutions with $K=Q=0$    are obtained:
\begin{eqnarray}
\left\{
\begin{array} {l}
\Pi _0 = - \Pi _1  \;   , \;
\omega = \pm \sqrt{ 4 / 3 + 8 \phi _1 - 4 W_1 ^2 + 8 \Pi _0} 
\; , \; \\
  \phi _1 = 1 / 12 \; , \; \gamma = 3 / 2 
\;  , \;  U _1 = -  \omega / 2 \; , \\  
V(\pi/2) =  V_0 =  \omega / 2 \; , \; V(0) = \omega. 
\end{array}
\right. 
\label{sol_rot2}
\end{eqnarray}

\begin{eqnarray}
\left\{
\begin{array} {l}
\Pi _0 = 0 \;  , \; \omega = \pm \sqrt{ 4 / 3 + 8 \phi _1 - 4 W_1 ^2} 
\; , \; \\
 \phi _1 = 1 / 12 +  \Pi _1 / 2 \;
 , \; \gamma = 1 \; , \;   U _1 = -  \omega / 2 \; , \\
V(\pi /2) =  V _0 =    \omega / 2 \; , \; V(0) = \omega.
\end{array}
\right.
\label{sol_rot3}
\end{eqnarray}

The polytropic index of the first solution (Eqs.~\ref{sol_rot2})
 is equal to $3/2$ and  the gravitational potential  is strongly
constrainted ($\phi _1=1/12$). The solution exists only if
$\Pi _ 0 > - 1 / 4 $ and   $W _1 < \sqrt{ 1/2 + 2 \Pi _0}$.

The polytropic index of the second solution (Eqs.~\ref{sol_rot3})
is equal to $1$. 
If $\Pi _1 = -1 /6$, the gravitational potential is isotropic ($\phi _1=0$).
 The solution exists only if  $W _1 < \sqrt{ 1/3 + 2 \phi _1}$.

\subsection{Aspherical pressure, rotation and magnetic field}
Two solutions with magnetic fields, rotation, aspherical pressure
are obtained:
\begin{eqnarray}
\left\{
\begin{array} {l}
\Pi _0 = - \Pi _1 \; , \; \\
 \Pi _0 = - 1/ 4 - Q ^2  +  W _1 ^2 /2  
-  ( \gamma -1 ) ( 3 \gamma -5) \omega ^2 / 2
 \; , \\
\omega  = \pm \sqrt{ { 2 \phi _1 - 1 / 6   \over (1 - \gamma) (3 - 2
\gamma) } } \; , \;
W _1 = 2 Q K { 4 - 3 \gamma   \over 3  - 2 \gamma } \; , \; \\
\; U _1 = (4 - 3 \gamma ) \omega \; , \\
V(\pi/2) = V _0 = (2- \gamma) \omega \; , \; V(0) = 2(\gamma -1) \omega.
\end{array}
\right. 
\label{sol_rot_mag2}
\end{eqnarray}

\begin{eqnarray}
\left\{
\begin{array} {l}
\Pi _0 = 0 \; , \; \\ 
\Pi _1 = - 1 / 2 - 2 Q ^2 + W _1 ^2 +   (3 \gamma -2)
(\gamma -2) \omega ^2 /4 \; , \\
\omega = \pm \sqrt{   8 \phi _1 + 4 /3  - 4 W _1 ^2 + 8 Q ^2  \over 
  \gamma  (2 - \gamma ) } \; , \; 
 W _1 = Q K { 3 \gamma - 4 \over \gamma -2}   \; , \; \\
U _1 =  (3 \gamma -4 )  \omega / 2   \; , \\
V(\pi/2) = V _0 = \gamma  \omega / 2 \; , \; 
V(0) = (2 - \gamma) \omega. 
\end{array}
\right.
\label{sol_rot_mag3}
\end{eqnarray}
The first solution (Eqs.~\ref{sol_rot_mag2}) is defined for $\phi _1 <
1/12$ and for $1 < \gamma < 3/2$ or $\phi _1 > 1/12$ and $\gamma <1$
or $\gamma > 3/2$. The second one
(Eqs.~\ref{sol_rot_mag3}) is defined if $W _1 ^ 2 < 1 /3 + 2 \phi _1 +
2 Q ^2 $. 
For both solutions $\phi _1$ can be equal to zero (isotropic
potential) and the rotation
parameter, $W _1$, is proportional to $Q K$.

As for Eqs.~\ref{sol_hyd2}, if $\gamma < 1$ and $\phi _1 > 1/12$, the
first solution presents an outflow at the pole.

\end{document}